\documentclass[iop, letterpaper]{emulateapj}
\usepackage{apjfonts}
\usepackage{rotating}
\usepackage{amsmath}

\newcommand{\teff}{\mbox{$T_{\rm eff}$}}
\newcommand{\logg}{\mbox{$\log g$}}
\newcommand{\feh}{\mbox{$\rm{[Fe/H]}$}}

\newcommand{\numax}{\mbox{$\nu_{\rm max}$}}
\newcommand{\dnu}{\mbox{$\Delta \nu$}}
\newcommand{\muHz}{\mbox{$\mu$Hz}}

\shorttitle{Probing the deep end of the Milky Way with \emph{Kepler}}
\shortauthors{MATHUR ET AL.}

\begin{document}

\title{Probing the deep end of the Milky Way with \emph{Kepler}:\\
Asteroseismic analysis of 854 faint Red Giants misclassified as Cool Dwarfs
}

\author{S. Mathur\altaffilmark{1}, R. A. Garc\'ia\altaffilmark{2}, D. Huber\altaffilmark{3,4,5}, C. Regulo\altaffilmark{6,7}, D. Stello\altaffilmark{3}, P. G. Beck\altaffilmark{2}, K. Houmani\altaffilmark{2}, and D. Salabert\altaffilmark{2}}

\altaffiltext{1}{Space Science Institute, 4750 Walnut street Suite\#205, Boulder, CO 80301, USA}
\altaffiltext{2}{Laboratoire AIM, CEA/DRF-CNRS-Universit\'e Paris Diderot; IRFU/SAp, Centre de Saclay, 91191 Gif-sur-Yvette Cedex, France}
\altaffiltext{3}{Sydney Institute for Astronomy (SIfA), School of Physics, University of Sydney, NSW 2006, Australia}
\altaffiltext{4}{SETI Institute, 189 Bernardo Avenue, Mountain View, CA 94043, USA}
\altaffiltext{5}{Stellar Astrophysics Centre, Department of Physics and Astronomy, 
Aarhus University, Ny Munkegade 120, DK-8000 Aarhus C, Denmark}
\altaffiltext{6}{Universidad de La Laguna, Dpto de Astrof\'isica, 38206, Tenerife, Spain}
\altaffiltext{7}{Instituto de Astrof\'\i sica de Canarias, 38205, La Laguna, Tenerife, Spain}

\begin{abstract}

Asteroseismology has proven to be an excellent tool to determine not only the global stellar properties with a good precision but also to infer  stellar structure, dynamics, and evolution for a large sample 
of {\it Kepler} stars. Prior to the launch of the mission the properties of Kepler targets were inferred from broadband photometry, leading to the Input Catalog \citep[KIC][]{2011AJ....142..112B}. The KIC was later revised in the {\it Kepler} Star Properties Catalog \citep{2014ApJS..211....2H}, based on literature values and an asteroseismic analysis of stars which were unclassified in the KIC. Here we present an asteroseismic analysis of 45,400 stars which were classified as dwarfs in the Kepler Star Properties Catalog. We found that around 2\% of the sample shows acoustic modes in the typical frequency range that put them in the red-giant category rather than cool dwarfs. We analyse the asteroseismic properties of these stars, derive their surface gravities, masses, and radii and present updated effective temperatures and distances. We show that the sample is significantly fainter than the previously known oscillating giants in the {\it Kepler} field, with the faintest stars reaching down to a {\it Kepler} magnitude, Kp\,$\sim$\,16. We demonstrate that 404 stars are at distances beyond 5 kpc and that the stars are significantly less massive than for the original {\it Kepler} red-giant sample, consistent with a population of distant halo giants. {  A comparison with a galactic population model shows that up to 40 stars might be genuine halo giants, which would increase the number of known asteroseismic halo stars by a factor of 4.} The detections presented here will provide a valuable sample for galactic archeology studies. 


\end{abstract}

\keywords{asteroseismology ---methods: data analysis---stars: evolution---stars: 
fundamental parameters---stars: oscillations}

\section{INTRODUCTION}

The {\it Kepler} mission has observed more than 196,000 stars \citep{2010ApJ...713L.109B} over four years. While the main objective of the mission was to search for exoplanets, a lot of effort has been undertaken to well characterize the properties of the stars, in particular the ones hosting planets. 

To prepare the mission, all potential targets were observed in broadband photometry, which was used to build the {\it Kepler} Input catalog \citep[KIC,][]{2011AJ....142..112B}. The choice of photometric data over spectroscopic classification was mainly due to the large number of targets. However, the KIC also made use of existing catalogs such as the Two Micron All Sky Survey, USNO-B1.0, Hipparcos, Tycho-2, and UCAC2. Since the beginning of the mission, this catalog has been widely used for characterizing the discovered exoplanet host stars or for targeting specific categories of stars for stellar physics studies. It was also compared to spectroscopic observations that became progressively available during the lifetime of the mission \citep[e.g.][]{2012MNRAS.423..122B,2012A&A...543A.160T} or to re-calibrated KIC photometry \citep{2012ApJS..199...30P}. These studies showed that, on average, solar-like stars and red giants have effective temperatures 3 to 5\% higher than the ones from the KIC. The analysis of \citet{2011ApJ...738L..28V}, based on asteroseimology, showed that for a sample of main-sequence stars the KIC overestimates the surface gravities by on average 0.23\,dex  for $\log$ g above 4.0\,dex and 0.30\,dex for $\log$ g below 4.0\,dex. More recently, \citet{2015AJ....150..148H} showed a similar trend for a subsample of red giants observed by the SDSS-III/APOGEE survey.

\citet{2014ApJS..211....2H} (hereafter H14) updated the stellar properties of the {\it Kepler} targets by fitting stellar parameters available in the literature to isochrones of a grid of stellar models. In that work, they also detected oscillations in almost 3,000 stars that were unclassified in the KIC, allowing them to classify these stars as red giants. The precise knowledge of the stellar properties, in particular the masses and radii of planet host stars, is important as it can have a dramatic impact on the estimation of the size of the planet detected with the transit method that is used by the {\it Kepler} mission. 

Having long-cadence data for all the stars, we can look for signatures of oscillations in the stars, which would help us classify them as dwarfs or red giants. Indeed, for most giants and some classical pulsators, the modes are visible below the long-cadence Nyquist frequency \citep{2013ApJ...767..127H}. Here, we analyzed {  GKM stars as described in Section 2 and classified as dwarfs} in the H14 catalog to confirm their status as dwarfs. In Section 2 we briefly describe the target selection and how the {\it Kepler} light curves were calibrated. Section 3 provides the seismic analysis of the stars where we detected oscillations in the {\it Kepler} data. We also derive revised effective temperatures and distances. In Section 4, we discuss the results of this sample and give our conclusions in Section 5.

\section{Star selection and photometric calibration}






\noindent From the H14 catalog, we selected GKM dwarfs according to the following criteria: $\log g \ge 4.25$ and $T_{\rm eff} \le 5500K$. {  This temperature cut ensures little pollution from F stars in our sample although we loose the hottest G-type stars. The log g cut ensures that we do not have subgiants in our sample.} We also added a few tens stars from the asteroseismic sample of solar-like stars from \citet{2011ApJ...732...54C} where red-giant-like oscillations were detected. These additional stars have $T_{\rm eff} > 5500K$. This led to a sample of 45,431 stars. We calibrated the long-cadence light curves following \citet{2011MNRAS.414L...6G} and Mathur et al. (in prep.) using all the {\it Kepler} quarters available. Given the way the {\it Kepler} mission operated some of these stars had a variety of observation lengths going from one month to almost 4 years.  We concatenated the quarters when the stars were observed for several quarters. We also applied the gap-filling technique as described in \citet{2015A&A...574A..18P} to reduce the effects of the observational window function \citep{2014A&A...568A..10G}. Gaps up to 783 data points (16 days) were filled. Then we visually checked the light curves and their associated power spectra. Out of the 45,431 stars, we flagged 1107 stars that clearly showed oscillation modes or a red-giant like excess of power in the power spectrum.  Most of the selected stars have effective temperatures above 4000\,K. This agrees with the fact that the dwarf-giant classification in the KIC is more reliable for $\teff < 4000\,K$ \citep{2012ApJ...753...90M}.

During our visual inspection of the power spectra, we divided the stars into three different categories for the seismic analysis described in Section 3.1: 1) stars with modes located below 10\,$\mu$Hz, 2) stars with modes between 10 and 200\,$\mu$Hz, and 3) stars with modes above 200\,$\mu$Hz. Because the modes of the first group are at very low frequency we applied a high-pass filter with a lower cut-off frequency. The last category corresponds to stars where we can observe modes that are above the Nyquist frequency and that could be reflected to lower frequency \citep{2014MNRAS.445..946C}. These stars were analyzed in a different way to avoid a wrong estimation of $\log g$. This led to a sample of {  stars to be studied of} 121 stars with modes below 10\,$\mu$Hz, 723 stars between 10 and 200\,$\mu$Hz, and 137 stars above 200\,$\mu$Hz. We also flagged 126 stars that had either lower signal-to-noise ratio (SNR) or a less clear oscillation pattern. {  In the following section we explain how to characterize these stars to confirm the existence of p-mode oscillations and thus define the final set of stars to be studied.}

In the final study sample, stars have been observed for different duration. {  We notice that among the 1107 stars, 125 stars have less than three quarters available, which means that they have a lower frequency resolution tin the power spectra.}




\section{Seismic characterization of the stars}

\subsection{Determination of the acoustic-mode global parameters}


We analyzed the light curves and power spectra of the 1107 stars selected above with the A2Z pipeline \citep[][]{2010A&A...511A..46M} and the SYD pipeline \citep{2009CoAst.160...74H} to determine the global parameters of the acoustic modes. In order to better see the modes below 10\,$\mu$Hz, we applied a high-pass filter with a boxcar width of 40 days corresponding to a cut-off frequency of 0.29\,$\mu$Hz so that the high-pass filter does not affect the very low-frequency modes. For the remaining stars, we used a boxcar width of 20 days (i.e. 0.58\,$\mu$Hz).


We applied an improved version of A2Z (hereafter A2Z+) to have a more precise value of the mean large frequency separation, $\Delta \nu$, which is the distance between modes of same degree and of consecutive orders. We computed a first estimation of $\dnu$ with the auto-correlation function \citep{2009arXiv0909.0782M}, determined the frequency of the highest $\ell\,=\,$0 mode close to $\numax$ ($\nu_{\ell=0}$), and then masked out the $\ell\,=\,1$ modes. We recomputed the mean large frequency separation by calculating the power spectrum of the masked power spectrum (PS2) between $\nu_{\rm l=0}-k\times \Delta\nu$ and $\nu_{\rm l=0}+k\times \Delta\nu$. In general we take $k\,=\,2$ but the value of $k$ can be different depending on the SNR and the frequency of maximum power, $\numax$, of the star. The A2Z+ uncertainties were computed with the weighted centroids method that depends on the frequency resolution in the PS2. Given that in most cases we take the same number of orders around $\nu_{\ell=0}$,  the relative uncertainties are generally very similar for the different stars. 

For the stars with modes above 200$\mu$Hz, we applied a slightly different methodology as the modes observed can result from the reflection of modes oscillating above the Nyquist frequency of 283.21\,$\mu$Hz \citep[see][for more details]{2014MNRAS.445..946C}. The reflected modes keep the properties of the real modes, in particular the mean large frequency separation. If the modes observed are actually oscillating above the Nyquist frequency the correspondence between the observed frequency of maximum power, $\nu_{\rm max}$, and $\Delta \nu$ will show clear disagreement with the $\dnu$-$\numax$ relation \citep{2009MNRAS.400L..80S}. To analyze these stars, we reconstructed the power spectrum by mirroring our power spectrum above  the Nyquist frequency. This allowed us to measure $\Delta \nu$ and $\nu_{\rm max}$ above and below $\nu_{\rm Nyq}$. Then we checked which $\nu_{\rm max}$ agreed with $\Delta \nu$ according to the seismic scaling relations \citep{1995A&A...293...87K,2009MNRAS.400L..80S}. {  We first confirmed 61 stars that have $\nu_{\rm max}$ between 200 and 285\,$\mu$Hz. After a visual inspection of the spectra, we also confirmed the detection of modes for 31 more stars with \numax\,$>$\, 285\,$\mu$Hz (three stars with modes above the Nyquist frequency according to our method and 28 stars with partially observed modes). This partial mode detection only provides a very rough estimate of \numax and it is not possible to provide a reliable estimation of a seismic \logg.} Finally, we checked visually the results  with \'echelle diagrams and adjusted $\dnu$ in order to have the straightest ridges for the $\ell\,=\,0$ modes and validated the seismic parameters for the A2Z+ pipeline. 

{  We found 875 stars with modes detected by either A2Z+ or SYD below the Nyquist frequency. Among them, eight stars have a low SNR and thus the detection is less clear, in particular the determination of the mean large separation. As we do see an excess of power, we list them in the same table without any value for $\Delta \nu$ and a special flag ('SNR').

We compared the results of the A2Z+ and SYD pipelines for 784 stars where both pipelines returned results. A more detailed description of the comparison between the two pipelines is given in the Appendix A.}





\subsection{Blending Effects}

\begin{figure}[htbp]
\begin{center}
\includegraphics[angle=90,width=9cm]{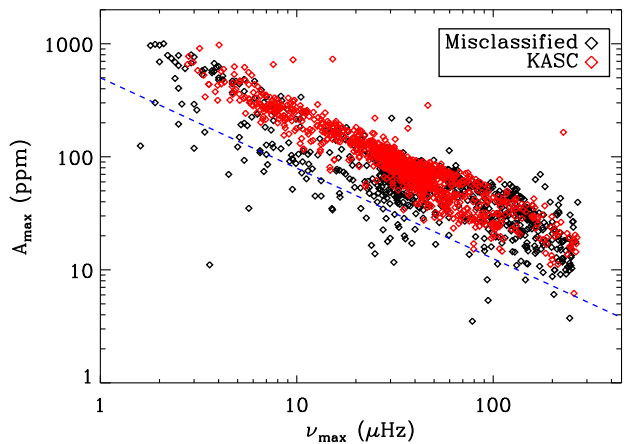}
\caption{Maximum amplitude per radial mode as a function of \numax for the confirmed detections in our sample ({  black} symbols) and for $\sim$1200 known red giants ({  red} symbols). The blue dashed line represents the $5.10^2\times\numax^{-0.8}$ below which we considered the stars with low $A_{\rm max}$.}
\label{Amax_numax}
\end{center}
\end{figure}

Some of the detected oscillations may be affected by stars polluting the photometric aperture of the giant star. To check that this was not the case, we fitted the background of the power spectra following  \citet{2011ApJ...741..119M} and \citet{2014A&A...570A..41K} and computed the maximum amplitudes per radial mode of the acoustic modes.  These parameters are related to the luminosity of the stars as already shown in previous works \citep{2011ApJ...743..143H,2011ApJ...737L..10S}. We plot in Figure~\ref{Amax_numax} the maximum amplitude per radial mode  as a function of $\nu_{\rm max}$ for the new classified red giants and compare them to the values calculated for a known sample of red giants studied within the {\it Kepler} Asteroseismic Consortium (KASC). {  These stars are red giants as listed in the KIC  and astrometric references stars \citep{2010ApJ...713L.109B}.} Based on the KASC sample, \citep{2011ApJ...743..143H} derived that $A_{\rm max} \propto \numax^{-0.8}$. From Figure~\ref{Amax_numax}, we can infer a lower limit of the relation ($5.10^2\times\numax^{-0.8}$) below which there are no KASC stars. We find 50 stars that have an abnormally low $A_{\rm max}$. This lower amplitude can be explained by the fact that if a red giant has a nearby companion, its light is going to be diluted by the nearby star causing a decrease of the  measured amplitude of the brightness variation. For this sample of stars, we checked the UKIRT images in the J band\footnote{\url{http://surveys.roe.ac.uk/wsa/}} within 2 arcmin of the target. We found that 36 stars out of the 50 stars with low $A_{\rm max}$ could indeed be polluted by a brighter nearby star. They are flagged in the Flag column of Table~\ref{tbl1} as 'PB' (for Possible Blend). {  For the remaining stars that have $A_{\rm max}$ in agreement with the KASC red giants}, the UKIRT images suggest that 62 stars have a nearby star that could pollute the main target. 


Some of the oscillations detected could be due to a known red giant leaking onto the pixels of the dwarf. We searched for known red giants (either with detected oscillations or classified as red giants in the H14 catalog) that were within a radius of 1 arcmin (15 pixels) from the misclassified stars. The comparison of the power spectra of nearby known red giants with the misclassified stars showed that  {  52 stars out of the 875 stars with modes below Nyquist} have the same power spectra and thus the mode detection results from pollution. We discarded these {  52} stars from our sample. There are three additional stars (KIC 4907817, 8246648, and 9788579) where the power spectra look similar but the length of data available is different for the two samples of stars without any overlapping quarters making it more difficult to compare them. {  This leads to a final sample of confirmed misclassified stars of  854 stars: 820 stars with \numax well below Nyquist, 31 near/superNyquist stars, and three stars that could result from pollution and that have the comment 'PP' (for Possible Pollution).}

\begin{figure}[htbp]
\begin{center}
\includegraphics[angle=90,width=9cm]{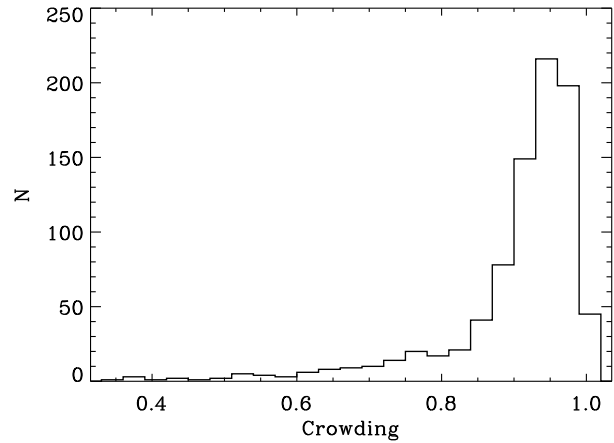}
\caption{Histogram of the crowding as described in Section 3.2 for the 854 misclassified red giants.}
\label{histo_crowd}
\end{center}
\end{figure}

{  In addition to the visual check of the J-band images for this subsample of stars, we also looked at the crowding of the whole sample of stars. The crowding is provided on the MAST\footnote{\url{https://archive.stsci.edu/kepler/}} and it is defined as the ratio of the target flux to the total flux in the optimal aperture. {  The optimal aperture is computed by the NASA Pre-search Data Conditioning \citep[for more details see][]{2010ApJ...713L..87J}. In this paper we define a crowding value of 1 when all the flux belongs to the main target. Since the satellite rolls every three months to keep its solar panels towards the Sun, the aperture and thus the crowding changes every quarter (as described in the NASA Kepler release notes). We computed the crowding as the median value of all the quarters.} The values of crowding for each target are listed in Tables~\ref{tbl1} and \ref{tbl3}.  {  Figure~\ref{histo_crowd} shows the distribution of the crowding for the 854 stars. Most of the stars have a crowding larger than 75\%}, meaning that most of the flux is coming from the main target and the probability that there is a nearby star polluting the light curve is low. {  Stars with a crowding value below 75\%} should be considered very cautiously. They have the comment 'CR' (for Crowding') in the last column of Table~\ref{tbl1}.}

Tables~\ref{tbl1} and \ref{tbl2} list the seismic parameters obtained by the A2Z+ and SYD pipelines respectively. Table~\ref{tbl3} gives the seismic parameters from A2Z+ for the near and super-Nyquist stars. 

One of these stars, KIC 7292836, could be a seismic binary where we see two regions with modes. After changing the aperture for this star, by using  a smaller number of pixels to produce the light curves, and inspecting the J-band image, we could not see a nearby star. We provide the seismic parameters of the two power excesses.




\begin{deluxetable*}{rcccccccrccc}
\tablewidth{0pt}
\tablecaption{Seismic and fundamental parameters of the misclassified red giants obtained for the A2Z+ pipeline\tablenotemark{+}}
\tablehead{\colhead{ KIC} & \colhead{$K_p$} & \colhead{$T_{\rm eff}$ (K) }& \colhead{$T^{*}_{\rm eff}$ (K)} & \colhead{$\Delta \nu$ ($\mu$Hz)} & \colhead{$\nu_{\rm max}$ ($\mu$Hz)} &  \colhead{$\log g$} & \colhead{M$_*$ ($M_{\odot}$)} & \colhead{R$_*$ ($R_{\odot}$)} & \colhead{d (kpc)} & \colhead{Crowding} & \colhead{Flag}}
\startdata
893233 &  11.4& 4204$^{+ 154}_{- 156}$& 4285$^{+  96}_{-  96}$& 
     1.03\,$\pm$\,     0.18&       6.1\,$\pm$\,      0.4& 1.67\,$\pm$\,0.03&
 1.51\,$\pm$\, 0.56 &   29.59\,$\pm$\,  7.58 &   1.66$^{+ 0.92}_{- 0.31}$ &  
0.98&  \\
1027110 &  12.1& 4415$^{+ 161}_{- 149}$& 4272$^{+  85}_{-  85}$& 
     1.13\,$\pm$\,     0.11&       6.6\,$\pm$\,      0.4& 1.71\,$\pm$\,0.03&
 1.31\,$\pm$\, 0.29 &   26.56\,$\pm$\,  4.01 &   2.06$^{+ 0.67}_{- 0.13}$ &  
0.98&  \\
1160684 &  14.9& 4137$^{+ 121}_{- 144}$& 4571$^{+  91}_{-  91}$& 
     3.59\,$\pm$\,     0.10&      27.5\,$\pm$\,      1.4& 2.34\,$\pm$\,0.02&
 1.03\,$\pm$\, 0.11 &   11.34\,$\pm$\,  0.75 &   3.76$^{+ 0.09}_{- 0.08}$ &  
0.91&  \\
1162220 &  11.2& 4284$^{+ 129}_{- 134}$& 4303$^{+  87}_{-  86}$& 
     1.68\,$\pm$\,     0.05&      11.1\,$\pm$\,      0.7& 1.94\,$\pm$\,0.03&
 1.29\,$\pm$\, 0.16 &   20.28\,$\pm$\,  1.56 &   1.21$^{+ 0.18}_{- 0.07}$ &  
0.99&  \\
1163114 &  12.7& 4211$^{+ 123}_{- 138}$& 4440$^{+  88}_{-  88}$& 
     1.91\,$\pm$\,     0.20&      14.2\,$\pm$\,      0.7& 2.05\,$\pm$\,0.02&
 1.70\,$\pm$\, 0.39 &   20.39\,$\pm$\,  3.20 &   2.63$^{+ 0.34}_{- 0.26}$ &  
0.97&  \\
1164356 &  14.1& 5225$^{+ 169}_{- 135}$& 5360$^{+ 178}_{- 125}$& 
     3.33\,$\pm$\,     0.09&      27.8\,$\pm$\,      1.8& 2.38\,$\pm$\,0.03&
 1.83\,$\pm$\, 0.24 &   14.43\,$\pm$\,  1.14 &   6.91$^{+ 0.21}_{- 0.25}$ &  
0.74&  \\
1164584 &  14.3& 4333$^{+ 128}_{- 133}$& 4546$^{+  90}_{-  90}$& 
     3.35\,$\pm$\,     0.12&      28.8\,$\pm$\,      1.4& 2.36\,$\pm$\,0.02&
 1.55\,$\pm$\, 0.18 &   13.60\,$\pm$\,  0.97 &   3.77$^{+ 0.29}_{- 0.54}$ &  
0.95&  \\
1292147 &  15.2& 4219$^{+ 121}_{- 145}$& 4764$^{+ 130}_{- 106}$& 
     8.12\,$\pm$\,     0.28&      87.2\,$\pm$\,      4.2& 2.85\,$\pm$\,0.02&
 1.34\,$\pm$\, 0.15 &    7.18\,$\pm$\,  0.51 &   3.20$^{+ 0.71}_{- 0.30}$ &  
0.91&  \\
1293587 &  14.9& 4914$^{+ 152}_{- 126}$& 4921$^{+ 149}_{-  98}$& 
    16.67\,$\pm$\,     0.17&     253.8\,$\pm$\,     23.8& 3.32\,$\pm$\,0.04&
 1.95\,$\pm$\, 0.33 &    5.04\,$\pm$\,  0.49 &   2.17$^{+ 0.18}_{- 0.12}$ &  
0.86&  \\
1429629 &  15.2& 4260$^{+ 129}_{- 146}$& 4697$^{+ 152}_{-  93}$& 
     6.28\,$\pm$\,     0.19&      63.0\,$\pm$\,      3.1& 2.71\,$\pm$\,0.02&
 1.38\,$\pm$\, 0.15 &    8.61\,$\pm$\,  0.59 &   3.94$^{+ 0.69}_{- 0.32}$ &  
0.94&  \\
1431059 &  13.1& 4660$^{+ 140}_{- 122}$& 4938$^{+ 101}_{-  98}$& 
    13.62\,$\pm$\,     0.47&     171.3\,$\pm$\,      7.6& 3.15\,$\pm$\,0.02&
 1.35\,$\pm$\, 0.14 &    5.10\,$\pm$\,  0.34 &   1.06$^{+ 0.07}_{- 0.07}$ &  
0.98&  \\
1431599 &  12.3& 4056$^{+ 100}_{- 115}$& 4007$^{+  80}_{-  80}$& 
     0.53\,$\pm$\,     0.28&       2.3\,$\pm$\,      0.2& 1.24\,$\pm$\,0.04&
 1.04\,$\pm$\, 1.11 &   40.75\,$\pm$\, 30.66 &   3.24$^{+ 0.56}_{- 0.21}$ &  
0.98&  \\
1432022 &  14.8& 5706$^{+ 162}_{- 184}$& 4726$^{+  94}_{-  94}$& 
     0.92\,$\pm$\,     0.01&       4.5\,$\pm$\,      0.3& 1.56\,$\pm$\,0.03&
 1.10\,$\pm$\, 0.13 &   28.73\,$\pm$\,  2.01 &  13.65$^{+ 2.83}_{- 0.77}$ &  
0.93&  PB \\
1576043 &  12.5& 4012$^{+ 120}_{- 149}$& 4176$^{+  83}_{-  83}$& 
     0.91\,$\pm$\,     0.03&       5.4\,$\pm$\,      0.4& 1.62\,$\pm$\,0.03&
 1.65\,$\pm$\, 0.24 &   33.13\,$\pm$\,  2.94 &   2.76$^{+ 0.46}_{- 0.27}$ &  
0.98&  \\
2010137 &  12.3& 4019$^{+ 123}_{- 158}$& 4253$^{+  85}_{-  85}$& 
     1.08\,$\pm$\,     0.15&       6.1\,$\pm$\,      0.5& 1.67\,$\pm$\,0.04&
 1.23\,$\pm$\, 0.39 &   26.81\,$\pm$\,  5.72 &   2.18$^{+ 1.19}_{- 0.22}$ &  
0.98&  \\
2010577 &  15.6& 4703$^{+ 181}_{- 159}$& 4773$^{+ 116}_{- 116}$& 
     3.90\,$\pm$\,     0.67&      30.0\,$\pm$\,      1.5& 2.39\,$\pm$\,0.02&
 1.03\,$\pm$\, 0.36 &   10.71\,$\pm$\,  2.66 &   6.78$^{+ 4.03}_{- 0.58}$ &  
0.34& CR \\
2011870 &  13.2& 3962$^{+ 116}_{- 161}$& 4118$^{+  83}_{-  83}$& 
     0.76\,$\pm$\,     0.01&       4.7\,$\pm$\,      0.2& 1.55\,$\pm$\,0.02&
 2.19\,$\pm$\, 0.18 &   41.05\,$\pm$\,  2.00 &   4.77$^{+ 0.42}_{- 0.25}$ &  
0.97&  \\
2014684 &  11.8& 4200$^{+ 124}_{- 138}$& 4293$^{+  85}_{-  85}$& 
     1.05\,$\pm$\,     0.11&       5.8\,$\pm$\,      0.4& 1.65\,$\pm$\,0.03&
 1.20\,$\pm$\, 0.29 &   27.10\,$\pm$\,  4.45 &   1.90$^{+ 0.81}_{- 0.19}$ &  
0.99&  \\
2014737 &  15.6& 5177$^{+ 214}_{- 160}$& 5441$^{+ 110}_{- 110}$& 
     3.93\,$\pm$\,     0.67&      27.4\,$\pm$\,      1.7& 2.38\,$\pm$\,0.03&
 0.92\,$\pm$\, 0.33 &   10.29\,$\pm$\,  2.57 &  22.54$^{+ 1.69}_{- 2.06}$ &  
0.35& CR \\
2159728 &  15.2& 4787$^{+ 189}_{- 157}$& 4936$^{+ 196}_{- 196}$& 
     7.33\,$\pm$\,     0.27&      78.0\,$\pm$\,      3.7& 2.81\,$\pm$\,0.02&
 1.52\,$\pm$\, 0.18 &    8.02\,$\pm$\,  0.61 &   4.71$^{+ 0.26}_{- 0.77}$ &  
0.92&  \\
2167183 &  13.7& 3987$^{+ 126}_{- 156}$& 3975$^{+  79}_{-  79}$& 
     0.55\,$\pm$\,     0.01&       2.5\,$\pm$\,      0.2& 1.27\,$\pm$\,0.04&
 1.14\,$\pm$\, 0.17 &   40.97\,$\pm$\,  3.49 &   5.40$^{+ 0.32}_{- 0.19}$ &  
0.92&  \\
2284679 &  11.4& 4895$^{+ 149}_{- 149}$& 5004$^{+ 126}_{- 113}$& 
    16.15\,$\pm$\,     0.55&     199.8\,$\pm$\,     11.5& 3.22\,$\pm$\,0.03&
 1.11\,$\pm$\, 0.14 &    4.26\,$\pm$\,  0.33 &   0.56$^{+ 0.04}_{- 0.04}$ &  
0.99&  \\
2297384 &  14.3& 4350$^{+ 125}_{- 137}$& 4625$^{+  92}_{-  92}$& 
     3.78\,$\pm$\,     0.12&      32.0\,$\pm$\,      2.0& 2.41\,$\pm$\,0.03&
 1.35\,$\pm$\, 0.17 &   11.98\,$\pm$\,  0.94 &   3.42$^{+ 0.34}_{- 0.14}$ &  
0.94&  \\
2297825 &  14.3& 4240$^{+ 123}_{- 140}$& 4668$^{+  93}_{-  93}$& 
     3.78\,$\pm$\,     0.13&      31.0\,$\pm$\,      1.7& 2.40\,$\pm$\,0.03&
 1.24\,$\pm$\, 0.15 &   11.65\,$\pm$\,  0.87 &   3.37$^{+ 0.19}_{- 0.16}$ &  
0.94&  \\
2308968 &  15.2& 4988$^{+ 156}_{- 162}$& 5097$^{+ 152}_{- 114}$& 
     5.02\,$\pm$\,     0.95&      33.5\,$\pm$\,      2.4& 2.45\,$\pm$\,0.03&
 0.57\,$\pm$\, 0.23 &    7.46\,$\pm$\,  2.07 &   9.23$^{+ 0.50}_{- 3.03}$ &  
0.91&  \\
2309524 &  12.1& 3839$^{+ 121}_{- 141}$& 4060$^{+ 102}_{-  92}$& 
     0.71\,$\pm$\,     0.15&       3.7\,$\pm$\,      0.2& 1.45\,$\pm$\,0.03&
 1.38\,$\pm$\, 0.60 &   36.77\,$\pm$\, 11.18 &   2.09$^{+ 0.56}_{- 0.12}$ &  
0.97&  \\
2422890 &  13.0& 3898$^{+ 125}_{- 152}$& 3899$^{+  88}_{-  88}$& 
     0.44\,$\pm$\,     0.00&       1.8\,$\pm$\,      0.1& 1.12\,$\pm$\,0.03&
 1.01\,$\pm$\, 0.10 &   45.65\,$\pm$\,  2.72 &   4.93$^{+ 0.20}_{- 0.25}$ &  
0.97&  \\
2436209 &  14.9& 4248$^{+ 128}_{- 139}$& 4716$^{+ 102}_{-  94}$& 
     5.74\,$\pm$\,     0.82&      57.2\,$\pm$\,      2.7& 2.67\,$\pm$\,0.02&
 1.49\,$\pm$\, 0.44 &    9.37\,$\pm$\,  1.95 &   4.31$^{+ 0.34}_{- 0.68}$ &  
0.89&  \\
2436332 &  14.3& 4187$^{+ 126}_{- 134}$& 4498$^{+  90}_{-  90}$& 
     3.39\,$\pm$\,     0.10&      28.7\,$\pm$\,      1.6& 2.36\,$\pm$\,0.03&
 1.44\,$\pm$\, 0.17 &   13.17\,$\pm$\,  0.94 &   3.71$^{+ 0.30}_{- 0.48}$ &  
0.92&  \\
2436540 &  14.9& 4386$^{+ 123}_{- 136}$& 4652$^{+ 116}_{-  93}$& 
     5.80\,$\pm$\,     0.15&      59.2\,$\pm$\,      2.8& 2.68\,$\pm$\,0.02&
 1.55\,$\pm$\, 0.16 &    9.44\,$\pm$\,  0.59 &   3.74$^{+ 0.45}_{- 0.26}$ &  
0.90&  \\
2436814 &  14.2& 4219$^{+ 126}_{- 136}$& 4488$^{+  93}_{-  89}$& 
     3.10\,$\pm$\,     0.11&      24.1\,$\pm$\,      1.3& 2.28\,$\pm$\,0.02&
 1.22\,$\pm$\, 0.15 &   13.21\,$\pm$\,  0.99 &   3.62$^{+ 0.47}_{- 0.35}$ &  
0.80&  \\
2436824 &  14.5& 4214$^{+ 126}_{- 139}$& 4429$^{+  88}_{-  88}$& 
     3.85\,$\pm$\,     0.12&      34.0\,$\pm$\,      1.6& 2.43\,$\pm$\,0.02&
 1.41\,$\pm$\, 0.15 &   12.00\,$\pm$\,  0.79 &   5.90$^{+ 0.33}_{- 0.56}$ &  
0.87&  \\
2437040 &  14.2& 4209$^{+ 118}_{- 144}$& 4490$^{+  89}_{-  89}$& 
     3.07\,$\pm$\,     0.08&      25.5\,$\pm$\,      1.2& 2.31\,$\pm$\,0.02&
 1.50\,$\pm$\, 0.15 &   14.25\,$\pm$\,  0.88 &   3.82$^{+ 0.22}_{- 0.34}$ &  
0.72&  \\
2437402 &  14.7& 4315$^{+ 124}_{- 139}$& 4712$^{+ 104}_{-  94}$& 
     4.81\,$\pm$\,     0.16&      46.1\,$\pm$\,      2.3& 2.57\,$\pm$\,0.02&
 1.58\,$\pm$\, 0.18 &   10.75\,$\pm$\,  0.76 &   4.11$^{+ 0.14}_{- 0.17}$ &  
0.69&  \\
2437507 &  14.1& 4156$^{+ 122}_{- 132}$& 4493$^{+  95}_{-  95}$& 
     2.62\,$\pm$\,     0.01&      20.6\,$\pm$\,      1.1& 2.21\,$\pm$\,0.02&
 1.49\,$\pm$\, 0.14 &   15.81\,$\pm$\,  0.88 &   4.00$^{+ 0.09}_{- 0.34}$ &  
0.94&  \\
2437539 &  14.6& 4178$^{+ 119}_{- 137}$& 4683$^{+ 104}_{-  93}$& 
     4.88\,$\pm$\,     0.17&      46.8\,$\pm$\,      2.3& 2.58\,$\pm$\,0.02&
 1.54\,$\pm$\, 0.18 &   10.57\,$\pm$\,  0.75 &   3.87$^{+ 0.18}_{- 0.47}$ &  
0.94&  \\
2437589 &  14.3& 4197$^{+ 119}_{- 138}$& 4729$^{+  94}_{-  94}$& 
     4.63\,$\pm$\,     0.16&      46.9\,$\pm$\,      2.4& 2.58\,$\pm$\,0.02&
 1.95\,$\pm$\, 0.22 &   11.83\,$\pm$\,  0.85 &   3.52$^{+ 0.21}_{- 0.13}$ &  
0.95&  \\
2437653 &  15.2& 4386$^{+ 130}_{- 129}$& 4818$^{+ 155}_{- 155}$& 
     7.03\,$\pm$\,     0.21&      74.1\,$\pm$\,      3.4& 2.78\,$\pm$\,0.02&
 1.49\,$\pm$\, 0.16 &    8.18\,$\pm$\,  0.54 &   4.25$^{+ 0.33}_{- 0.71}$ &  
0.86&  \\
2437698 &  14.2& 4269$^{+ 123}_{- 145}$& 4660$^{+  95}_{-  95}$& 
     3.74\,$\pm$\,     0.12&      31.8\,$\pm$\,      2.1& 2.41\,$\pm$\,0.03&
 1.39\,$\pm$\, 0.19 &   12.20\,$\pm$\,  0.99 &   3.37$^{+ 0.41}_{- 0.15}$ &  
0.94&  \\
2437805 &  14.2& 4178$^{+ 117}_{- 139}$& 4673$^{+  95}_{-  95}$& 
     3.85\,$\pm$\,     0.50&      31.8\,$\pm$\,      1.8& 2.41\,$\pm$\,0.03&
 1.25\,$\pm$\, 0.35 &   11.53\,$\pm$\,  2.22 &   3.32$^{+ 0.66}_{- 0.22}$ &  
0.93&  \\
2437816 &  14.0& 4146$^{+ 125}_{- 131}$& 4419$^{+  88}_{-  88}$& 
     2.38\,$\pm$\,     0.07&      17.9\,$\pm$\,      0.9& 2.15\,$\pm$\,0.02&
 1.40\,$\pm$\, 0.15 &   16.52\,$\pm$\,  1.10 &   3.79$^{+ 0.41}_{- 0.50}$ &  
0.91&  \\
2438038 &  15.0& 4315$^{+ 123}_{- 142}$& 4663$^{+ 103}_{-  93}$& 
     6.15\,$\pm$\,     0.19&      62.9\,$\pm$\,      3.0& 2.71\,$\pm$\,0.02&
 1.48\,$\pm$\, 0.16 &    8.93\,$\pm$\,  0.59 &   3.70$^{+ 0.34}_{- 0.27}$ &  
0.91&  \\
2438094 &  15.3& 4356$^{+ 124}_{- 138}$& 4369$^{+  87}_{-  87}$& 
     1.58\,$\pm$\,     0.04&       9.9\,$\pm$\,      0.7& 1.89\,$\pm$\,0.03&
 1.20\,$\pm$\, 0.16 &   20.61\,$\pm$\,  1.66 &   9.02$^{+ 0.82}_{- 0.65}$ &  
0.82&  \\
2441711 &  14.8& 4777$^{+ 187}_{- 163}$& 4696$^{+  93}_{-  93}$& 
     4.00\,$\pm$\,     0.11&      33.3\,$\pm$\,      3.6& 2.43\,$\pm$\,0.05&
 1.24\,$\pm$\, 0.24 &   11.21\,$\pm$\,  1.30 &   4.91$^{+ 0.24}_{- 0.19}$ &  
0.94&  \\
2448727 &  15.3& 4497$^{+ 168}_{- 156}$& 4645$^{+  92}_{-  92}$& 
     4.38\,$\pm$\,     0.13&      38.2\,$\pm$\,      2.3& 2.49\,$\pm$\,0.03&
 1.28\,$\pm$\, 0.16 &   10.67\,$\pm$\,  0.80 &   5.53$^{+ 0.20}_{- 0.59}$ &  
0.92&  \\
2556387 &  11.1& 3445$^{+ 117}_{- 113}$& 4464$^{+  89}_{-  89}$& 
     2.21\,$\pm$\,     0.06&      19.5\,$\pm$\,      1.0& 2.19\,$\pm$\,0.02&
 2.47\,$\pm$\, 0.26 &   20.97\,$\pm$\,  1.38 &   1.44$^{+ 0.10}_{- 0.08}$ &  
0.99&  \\
2569360 &  14.2& 4187$^{+ 122}_{- 138}$& 4424$^{+  90}_{-  88}$& 
     2.75\,$\pm$\,     0.10&      21.3\,$\pm$\,      1.1& 2.22\,$\pm$\,0.02&
 1.33\,$\pm$\, 0.16 &   14.73\,$\pm$\,  1.09 &   3.71$^{+ 0.45}_{- 0.45}$ &  
0.94&  \\
2570094 &  15.0& 4386$^{+ 126}_{- 135}$& 4799$^{+ 147}_{- 132}$& 
     6.49\,$\pm$\,     0.22&      68.4\,$\pm$\,      3.2& 2.75\,$\pm$\,0.02&
 1.60\,$\pm$\, 0.18 &    8.84\,$\pm$\,  0.62 &   4.19$^{+ 0.40}_{- 0.65}$ &  
0.83&  \\
2570518 &  14.7& 4399$^{+ 130}_{- 127}$& 4566$^{+  91}_{-  91}$& 
     4.94\,$\pm$\,     0.15&      46.1\,$\pm$\,      2.7& 2.57\,$\pm$\,0.03&
 1.35\,$\pm$\, 0.16 &   10.04\,$\pm$\,  0.74 &   3.57$^{+ 0.31}_{- 0.21}$ &  
0.83&  \\
2571093 &  12.7& 3815$^{+ 114}_{- 139}$& 3900$^{+  78}_{-  78}$& 
     0.48\,$\pm$\,     0.48&       2.1\,$\pm$\,      0.1& 1.19\,$\pm$\,0.02&
 1.13\,$\pm$\, 2.27 &   44.75\,$\pm$\, 63.33 &   3.21$^{+ 0.39}_{- 0.19}$ &  
0.97&  \\
2578811 &  12.0& 3685$^{+ 110}_{- 139}$& 3912$^{+  80}_{-  78}$& 
     0.35\,$\pm$\,     0.04&       1.6\,$\pm$\,      0.1& 1.07\,$\pm$\,0.03&
 1.78\,$\pm$\, 0.45 &   64.22\,$\pm$\, 11.17 &   2.89$^{+ 1.04}_{- 0.30}$ &  
0.98&  \\
2585447 &  14.8& 5403$^{+ 168}_{- 137}$& 5305$^{+ 388}_{- 388}$& 
     1.42\,$\pm$\,     0.05&       8.3\,$\pm$\,      0.5& 1.85\,$\pm$\,0.04&
 1.45\,$\pm$\, 0.22 &   23.57\,$\pm$\,  2.21 &  19.55$^{+ 0.80}_{- 8.41}$ &  
0.56& CR \\
2708445 &  15.0& 4182$^{+ 120}_{- 135}$& 4584$^{+  91}_{-  91}$& 
     6.11\,$\pm$\,     0.18&      60.5\,$\pm$\,      3.0& 2.69\,$\pm$\,0.02&
 1.32\,$\pm$\, 0.14 &    8.63\,$\pm$\,  0.57 &   3.57$^{+ 0.22}_{- 0.25}$ &  
0.90&  \\
2713304 &  15.9& 3989$^{+  50}_{-  50}$& 4381$^{+ 108}_{-  97}$& 
     1.23\,$\pm$\,     0.20&       6.9\,$\pm$\,      0.6& 1.73\,$\pm$\,0.04&
 1.11\,$\pm$\, 0.40 &   23.73\,$\pm$\,  5.85 &  12.21$^{+ 7.20}_{- 1.35}$ &  
0.37& CR \\
2715419 &  15.1& 5151$^{+ 193}_{- 155}$& 5026$^{+ 100}_{- 100}$& 
     1.88\,$\pm$\,     0.01&       9.5\,$\pm$\,      0.7& 1.90\,$\pm$\,0.03&
 0.65\,$\pm$\, 0.09 &   14.98\,$\pm$\,  1.13 &   8.76$^{+ 0.39}_{- 0.18}$ &  
0.88&  \\
2721453 &  15.1& 4819$^{+ 154}_{- 134}$& 4882$^{+  97}_{-  97}$& 
    15.44\,$\pm$\,     0.05&     214.1\,$\pm$\,     35.3& 3.25\,$\pm$\,0.08&
 1.57\,$\pm$\, 0.45 &    4.93\,$\pm$\,  0.82 &   2.25$^{+ 0.15}_{- 0.13}$ &  
0.92&  \\
2845610 &   9.3& 5095$^{+ 151}_{- 129}$& 5051$^{+ 101}_{- 101}$& 
     6.97\,$\pm$\,     0.14&      90.1\,$\pm$\,      5.2& 2.88\,$\pm$\,0.03&
 2.97\,$\pm$\, 0.33 &   10.36\,$\pm$\,  0.68 &   0.46$^{+ 0.01}_{- 0.02}$ &  
1.00&  \\
2846051 &  10.2& 4203$^{+ 100}_{- 121}$& 4297$^{+  88}_{-  88}$& 
     1.15\,$\pm$\,     0.04&       6.8\,$\pm$\,      0.4& 1.72\,$\pm$\,0.03&
 1.35\,$\pm$\, 0.17 &   26.50\,$\pm$\,  2.07 &   0.96$^{+ 0.19}_{- 0.07}$ &  
1.00&  \\
...
\enddata
\tablenotetext{+}{\footnotesize $T_{\rm eff}$, $T_{\rm eff}*$, \logg, $M_{*}$, $R_*$, and d are respectively the values of effective temperature from H14, the revised effective temperature from \S~3.3, surface gravity, mass, radius from scaling relations and distance from isochrone fitting. The Flags 'PB',  'CR', and 'PP' stand respectively for 'Possible Blend', 'Crowding', and 'Possible Pollution'.}
\label{tbl1}
\end{deluxetable*}%

\begin{deluxetable*}{rccccccc}
\tablewidth{0pt}
\tablecaption{Seismic and fundamental parameters of the misclassified red giants obtained for the SYD pipeline\tablenotemark{+}}
\tablehead{\colhead{ KIC} &  \colhead{$T^{*}_{\rm eff}$ (K)} & \colhead{$\Delta \nu$ ($\mu$Hz)} & \colhead{$\nu_{\rm max}$ ($\mu$Hz)} &  \colhead{$\log g$} & \colhead{M$_*$ ($M_{\odot}$)} & \colhead{R$_*$ ($R_{\odot}$)} &  \colhead{d (kpc)} }
\startdata
893233 &  4276$^{+  89}_{- 156}$&      1.17\,$\pm$\,     0.02& 
      6.1\,$\pm$\,      0.1& 1.67\,$\pm$\,0.01& 0.90\,$\pm$\, 0.04 &  
 22.87\,$\pm$\,  0.71 &   1.44$^{+ 0.06}_{- 0.03}$ \\
1027110 &  4241$^{+  87}_{- 149}$&      1.15\,$\pm$\,     0.01& 
      6.7\,$\pm$\,      0.2& 1.71\,$\pm$\,0.02& 1.24\,$\pm$\, 0.09 &  
 25.79\,$\pm$\,  1.12 &   2.13$^{+ 0.14}_{- 0.12}$ \\
1160684 &  4555$^{+  91}_{- 144}$&      3.56\,$\pm$\,     0.05& 
     27.3\,$\pm$\,      0.6& 2.34\,$\pm$\,0.01& 1.04\,$\pm$\, 0.06 &  
 11.43\,$\pm$\,  0.39 &   3.77$^{+ 0.09}_{- 0.08}$ \\
1162220 &  4278$^{+ 106}_{- 134}$&      1.67\,$\pm$\,     0.01& 
     11.0\,$\pm$\,      0.4& 1.93\,$\pm$\,0.02& 1.28\,$\pm$\, 0.09 &  
 20.30\,$\pm$\,  0.80 &   1.22$^{+ 0.10}_{- 0.05}$ \\
1163114 &  4419$^{+  90}_{- 138}$&      1.89\,$\pm$\,     0.01& 
     14.3\,$\pm$\,      0.5& 2.05\,$\pm$\,0.02& 1.80\,$\pm$\, 0.12 &  
 20.96\,$\pm$\,  0.81 &   2.61$^{+ 0.14}_{- 0.10}$ \\
1164356 &  5471$^{+ 269}_{- 135}$&      3.33\,$\pm$\,     0.01& 
     27.7\,$\pm$\,      0.6& 2.38\,$\pm$\,0.02& 1.85\,$\pm$\, 0.13 &  
 14.47\,$\pm$\,  0.60 &   6.15$^{+ 0.81}_{- 0.68}$ \\
1164584 &  4554$^{+  95}_{- 133}$&      3.36\,$\pm$\,     0.01& 
     28.4\,$\pm$\,      0.5& 2.36\,$\pm$\,0.01& 1.49\,$\pm$\, 0.06 &  
 13.40\,$\pm$\,  0.30 &   3.76$^{+ 0.08}_{- 0.21}$ \\
1292147 &  4701$^{+  94}_{- 145}$&      8.12\,$\pm$\,     0.02& 
     88.0\,$\pm$\,      0.9& 2.85\,$\pm$\,0.01& 1.35\,$\pm$\, 0.04 &  
  7.20\,$\pm$\,  0.13 &   3.12$^{+ 0.13}_{- 0.11}$ \\
1429629 &  4664$^{+  93}_{- 146}$&      6.27\,$\pm$\,     0.01& 
     62.7\,$\pm$\,      0.7& 2.71\,$\pm$\,0.01& 1.36\,$\pm$\, 0.04 &  
  8.57\,$\pm$\,  0.16 &   2.12$^{+ 0.12}_{- 0.14}$ \\
1431059 &  4912$^{+  98}_{- 122}$&     13.64\,$\pm$\,     0.02& 
    169.5\,$\pm$\,      0.9& 3.15\,$\pm$\,0.01& 1.29\,$\pm$\, 0.03 &  
  5.02\,$\pm$\,  0.08 &   3.84$^{+ 0.18}_{- 0.10}$ \\
1431599 &  3989$^{+  79}_{- 115}$&      0.53\,$\pm$\,     0.01& 
      2.3\,$\pm$\,      0.1& 1.24\,$\pm$\,0.01& 1.01\,$\pm$\, 0.07 &  
 40.17\,$\pm$\,  1.63 &   1.05$^{+ 0.03}_{- 0.03}$ \\
1432022 &  4723$^{+  94}_{- 184}$&      0.90\,$\pm$\,     0.01& 
      4.4\,$\pm$\,      0.2& 1.55\,$\pm$\,0.02& 1.08\,$\pm$\, 0.11 &  
 28.98\,$\pm$\,  1.71 &   3.24$^{+ 0.15}_{- 0.15}$ \\
1576043 &  4168$^{+  83}_{- 149}$&      0.93\,$\pm$\,     0.01& 
      5.3\,$\pm$\,      0.3& 1.61\,$\pm$\,0.03& 1.44\,$\pm$\, 0.17 &  
 31.31\,$\pm$\,  2.11 &  13.71$^{+ 1.43}_{- 0.71}$ \\
2010137 &  4203$^{+  95}_{- 158}$&      1.08\,$\pm$\,     0.01& 
      6.0\,$\pm$\,      0.1& 1.66\,$\pm$\,0.01& 1.12\,$\pm$\, 0.05 &  
 25.96\,$\pm$\,  0.71 &   2.65$^{+ 0.42}_{- 0.17}$ \\
2010577 &  4749$^{+  95}_{- 159}$&      3.89\,$\pm$\,     0.05& 
     29.9\,$\pm$\,      0.6& 2.39\,$\pm$\,0.01& 1.02\,$\pm$\, 0.05 &  
 10.71\,$\pm$\,  0.33 &   2.18$^{+ 0.15}_{- 0.08}$ \\
2011870 &  4102$^{+  84}_{- 161}$&      0.77\,$\pm$\,     0.01& 
      4.6\,$\pm$\,      0.1& 1.54\,$\pm$\,0.01& 1.89\,$\pm$\, 0.08 &  
 38.54\,$\pm$\,  0.99 &   6.68$^{+ 0.30}_{- 0.30}$ \\
2014684 &  4282$^{+  85}_{- 138}$&      1.09\,$\pm$\,     0.02& 
      5.8\,$\pm$\,      0.1& 1.65\,$\pm$\,0.01& 1.03\,$\pm$\, 0.06 &  
 25.02\,$\pm$\,  0.91 &   4.64$^{+ 0.14}_{- 0.19}$ \\
2014737 &  5306$^{+ 106}_{- 160}$&      3.79\,$\pm$\,     0.10& 
     27.9\,$\pm$\,      0.8& 2.38\,$\pm$\,0.01& 1.10\,$\pm$\, 0.09 &  
 11.17\,$\pm$\,  0.56 &   1.84$^{+ 0.08}_{- 0.08}$ \\
2159728 &  4735$^{+  94}_{- 157}$&      7.32\,$\pm$\,     0.02& 
     78.0\,$\pm$\,      0.7& 2.80\,$\pm$\,0.01& 1.43\,$\pm$\, 0.04 &  
  7.87\,$\pm$\,  0.14 &  13.11$^{+ 1.40}_{- 0.93}$ \\
2167183 &  3993$^{+  79}_{- 156}$&      0.57\,$\pm$\,     0.02& 
      2.5\,$\pm$\,      0.0& 1.27\,$\pm$\,0.01& 1.03\,$\pm$\, 0.08 &  
 38.87\,$\pm$\,  2.00 &   3.91$^{+ 0.17}_{- 0.17}$ \\
2284679 &  4946$^{+ 143}_{- 149}$&     16.15\,$\pm$\,     0.02& 
    197.8\,$\pm$\,      1.0& 3.22\,$\pm$\,0.01& 1.06\,$\pm$\, 0.04 &  
  4.20\,$\pm$\,  0.09 &   5.23$^{+ 0.29}_{- 0.20}$ \\
2297384 &  4597$^{+  92}_{- 137}$&      3.79\,$\pm$\,     0.03& 
     31.2\,$\pm$\,      1.1& 2.40\,$\pm$\,0.02& 1.22\,$\pm$\, 0.08 &  
 11.57\,$\pm$\,  0.46 &   0.52$^{+ 0.03}_{- 0.02}$ \\
2297825 &  4654$^{+  93}_{- 140}$&      3.82\,$\pm$\,     0.03& 
     30.5\,$\pm$\,      0.8& 2.39\,$\pm$\,0.01& 1.13\,$\pm$\, 0.06 &  
 11.20\,$\pm$\,  0.36 &   3.42$^{+ 0.12}_{- 0.10}$ \\
2308968 &  5443$^{+ 108}_{- 162}$&      5.03\,$\pm$\,     0.11& 
     34.1\,$\pm$\,      0.6& 2.47\,$\pm$\,0.01& 0.66\,$\pm$\, 0.04 &  
  7.80\,$\pm$\,  0.29 &   3.40$^{+ 0.08}_{- 0.12}$ \\
2309524 &  4021$^{+ 100}_{- 141}$&      0.71\,$\pm$\,     0.01& 
      3.8\,$\pm$\,      0.1& 1.45\,$\pm$\,0.01& 1.43\,$\pm$\, 0.09 &  
 37.10\,$\pm$\,  1.44 &   5.19$^{+ 0.10}_{- 0.10}$ \\
2422890 &  3874$^{+  77}_{- 152}$&      0.42\,$\pm$\,     0.04& 
      1.7\,$\pm$\,      0.0& 1.09\,$\pm$\,0.01& 0.97\,$\pm$\, 0.18 &  
 46.27\,$\pm$\,  6.04 &   2.29$^{+ 0.27}_{- 0.15}$ \\
2436209 &  4668$^{+  93}_{- 139}$&      5.74\,$\pm$\,     0.01& 
     57.2\,$\pm$\,      0.6& 2.66\,$\pm$\,0.01& 1.46\,$\pm$\, 0.04 &  
  9.31\,$\pm$\,  0.16 &   4.96$^{+ 0.41}_{- 0.20}$ \\
2436332 &  4498$^{+  90}_{- 134}$&      3.40\,$\pm$\,     0.01& 
     28.5\,$\pm$\,      0.8& 2.35\,$\pm$\,0.01& 1.40\,$\pm$\, 0.07 &  
 13.02\,$\pm$\,  0.40 &   3.76$^{+ 0.13}_{- 0.17}$ \\
2436540 &  4652$^{+  93}_{- 136}$&      5.80\,$\pm$\,     0.01& 
     58.3\,$\pm$\,      0.8& 2.67\,$\pm$\,0.01& 1.49\,$\pm$\, 0.05 &  
  9.31\,$\pm$\,  0.19 &   3.67$^{+ 0.27}_{- 0.13}$ \\
2436814 &  4514$^{+  90}_{- 136}$&      3.13\,$\pm$\,     0.01& 
     24.4\,$\pm$\,      0.4& 2.29\,$\pm$\,0.01& 1.23\,$\pm$\, 0.05 &  
 13.17\,$\pm$\,  0.30 &   3.74$^{+ 0.18}_{- 0.16}$ \\
2436824 &  4417$^{+  88}_{- 139}$&      3.85\,$\pm$\,     0.01& 
     34.0\,$\pm$\,      0.3& 2.43\,$\pm$\,0.01& 1.40\,$\pm$\, 0.04 &  
 11.99\,$\pm$\,  0.21 &   3.61$^{+ 0.13}_{- 0.18}$ \\
2437040 &  4491$^{+  89}_{- 144}$&      3.08\,$\pm$\,     0.01& 
     25.5\,$\pm$\,      0.5& 2.31\,$\pm$\,0.01& 1.48\,$\pm$\, 0.06 &  
 14.18\,$\pm$\,  0.34 &   5.77$^{+ 0.32}_{- 0.29}$ \\
2437402 &  4688$^{+ 221}_{- 139}$&      4.81\,$\pm$\,     0.01& 
     45.7\,$\pm$\,      0.9& 2.57\,$\pm$\,0.02& 1.53\,$\pm$\, 0.10 &  
 10.65\,$\pm$\,  0.42 &   3.78$^{+ 0.09}_{- 0.23}$ \\
2437507 &  4486$^{+  89}_{- 132}$&      2.60\,$\pm$\,     0.01& 
     20.5\,$\pm$\,      0.4& 2.21\,$\pm$\,0.01& 1.50\,$\pm$\, 0.06 &  
 15.91\,$\pm$\,  0.39 &   4.06$^{+ 0.21}_{- 0.27}$ \\
2437539 &  4715$^{+ 194}_{- 137}$&      4.90\,$\pm$\,     0.01& 
     46.8\,$\pm$\,      0.6& 2.58\,$\pm$\,0.01& 1.53\,$\pm$\, 0.08 &  
 10.52\,$\pm$\,  0.34 &   3.95$^{+ 0.13}_{- 0.26}$ \\
2437589 &  4720$^{+  94}_{- 138}$&      4.61\,$\pm$\,     0.01& 
     46.6\,$\pm$\,      1.0& 2.58\,$\pm$\,0.01& 1.94\,$\pm$\, 0.09 &  
 11.86\,$\pm$\,  0.30 &   3.73$^{+ 0.30}_{- 0.18}$ \\
2437653 &  4688$^{+ 115}_{- 129}$&      7.01\,$\pm$\,     0.01& 
     73.8\,$\pm$\,      0.7& 2.78\,$\pm$\,0.01& 1.43\,$\pm$\, 0.05 &  
  8.09\,$\pm$\,  0.16 &   3.66$^{+ 0.11}_{- 0.18}$ \\
2437698 &  4643$^{+  93}_{- 145}$&      3.76\,$\pm$\,     0.02& 
     30.8\,$\pm$\,      1.6& 2.40\,$\pm$\,0.02& 1.24\,$\pm$\, 0.11 &  
 11.70\,$\pm$\,  0.62 &   3.68$^{+ 0.22}_{- 0.15}$ \\
2437805 &  4601$^{+  94}_{- 139}$&      3.79\,$\pm$\,     0.04& 
     31.7\,$\pm$\,      0.9& 2.41\,$\pm$\,0.01& 1.29\,$\pm$\, 0.08 &  
 11.79\,$\pm$\,  0.43 &   3.35$^{+ 0.19}_{- 0.13}$ \\
2437816 &  4421$^{+ 106}_{- 131}$&      2.35\,$\pm$\,     0.01& 
     17.7\,$\pm$\,      0.3& 2.14\,$\pm$\,0.01& 1.44\,$\pm$\, 0.06 &  
 16.83\,$\pm$\,  0.41 &   3.33$^{+ 0.17}_{- 0.08}$ \\
2438038 &  4647$^{+  92}_{- 142}$&      6.14\,$\pm$\,     0.01& 
     62.6\,$\pm$\,      0.7& 2.70\,$\pm$\,0.01& 1.46\,$\pm$\, 0.05 &  
  8.89\,$\pm$\,  0.16 &   3.76$^{+ 0.15}_{- 0.17}$ \\
2438094 &  4373$^{+  88}_{- 138}$&      1.58\,$\pm$\,     0.01& 
     10.0\,$\pm$\,      0.5& 1.89\,$\pm$\,0.02& 1.23\,$\pm$\, 0.11 &  
 20.76\,$\pm$\,  1.10 &   3.73$^{+ 0.10}_{- 0.15}$ \\
2441711 &  4678$^{+  93}_{- 163}$&      4.00\,$\pm$\,     0.04& 
     35.4\,$\pm$\,      1.2& 2.46\,$\pm$\,0.02& 1.47\,$\pm$\, 0.10 &  
 11.88\,$\pm$\,  0.46 &   9.00$^{+ 0.32}_{- 0.64}$ \\
2448727 &  4602$^{+ 126}_{- 156}$&      4.38\,$\pm$\,     0.01& 
     38.5\,$\pm$\,      0.8& 2.49\,$\pm$\,0.01& 1.28\,$\pm$\, 0.06 &  
 10.68\,$\pm$\,  0.31 &   4.95$^{+ 0.30}_{- 0.15}$ \\
2556387 &  4465$^{+  89}_{- 113}$&      2.23\,$\pm$\,     0.02& 
     19.4\,$\pm$\,      0.9& 2.19\,$\pm$\,0.02& 2.35\,$\pm$\, 0.20 &  
 20.48\,$\pm$\,  1.02 &   5.04$^{+ 0.57}_{- 0.16}$ \\
2569360 &  4393$^{+ 111}_{- 138}$&      2.75\,$\pm$\,     0.01& 
     21.5\,$\pm$\,      0.4& 2.23\,$\pm$\,0.01& 1.36\,$\pm$\, 0.06 &  
 14.85\,$\pm$\,  0.41 &   1.44$^{+ 0.10}_{- 0.05}$ \\
2570094 &  4722$^{+ 129}_{- 135}$&      6.50\,$\pm$\,     0.02& 
     67.8\,$\pm$\,      0.8& 2.74\,$\pm$\,0.01& 1.51\,$\pm$\, 0.06 &  
  8.66\,$\pm$\,  0.20 &   3.68$^{+ 0.14}_{- 0.16}$ \\
2570518 &  4563$^{+  91}_{- 127}$&      4.93\,$\pm$\,     0.01& 
     45.9\,$\pm$\,      0.6& 2.56\,$\pm$\,0.01& 1.34\,$\pm$\, 0.04 &  
 10.01\,$\pm$\,  0.19 &   3.79$^{+ 0.26}_{- 0.21}$ \\
2571093 &  3905$^{+  86}_{- 139}$&      0.45\,$\pm$\,     0.05& 
      2.1\,$\pm$\,      0.0& 1.19\,$\pm$\,0.01& 1.44\,$\pm$\, 0.32 &  
 50.61\,$\pm$\,  8.02 &   3.62$^{+ 0.13}_{- 0.09}$ \\
2578811 &  3910$^{+  78}_{- 139}$&      0.36\,$\pm$\,     0.02& 
      1.6\,$\pm$\,      0.2& 1.08\,$\pm$\,0.06& 1.71\,$\pm$\, 0.41 &  
 62.61\,$\pm$\,  9.00 &   3.25$^{+ 0.34}_{- 0.19}$ \\
2585447 &  4888$^{+ 432}_{- 137}$&      1.43\,$\pm$\,     0.02& 
      8.2\,$\pm$\,      0.6& 1.83\,$\pm$\,0.04& 1.21\,$\pm$\, 0.20 &  
 22.13\,$\pm$\,  2.17 &   2.99$^{+ 0.68}_{- 0.25}$ \\
2708445 &  4576$^{+  91}_{- 135}$&      6.09\,$\pm$\,     0.01& 
     60.2\,$\pm$\,      0.6& 2.68\,$\pm$\,0.01& 1.31\,$\pm$\, 0.04 &  
  8.63\,$\pm$\,  0.15 &  10.50$^{+ 8.06}_{- 1.42}$ \\
2713304 &  4380$^{+ 109}_{-  50}$&      1.14\,$\pm$\,     0.02& 
      6.6\,$\pm$\,      1.1& 1.72\,$\pm$\,0.08& 1.33\,$\pm$\, 0.39 &  
 26.53\,$\pm$\,  4.50 &   3.63$^{+ 0.11}_{- 0.13}$ \\
2715419 &  4999$^{+ 100}_{- 155}$&      1.84\,$\pm$\,     0.03& 
      9.5\,$\pm$\,      0.3& 1.90\,$\pm$\,0.02& 0.70\,$\pm$\, 0.05 &  
 15.54\,$\pm$\,  0.64 &  13.34$^{+ 4.44}_{- 1.48}$ \\
2845610 &  5055$^{+ 101}_{- 129}$&      7.13\,$\pm$\,     0.03& 
     93.6\,$\pm$\,      6.9& 2.90\,$\pm$\,0.03& 3.04\,$\pm$\, 0.40 &  
 10.29\,$\pm$\,  0.77 &   8.86$^{+ 0.25}_{- 0.18}$ \\
2846051 &  4322$^{+  86}_{- 121}$&      1.10\,$\pm$\,     0.01& 
      6.8\,$\pm$\,      0.2& 1.73\,$\pm$\,0.01& 1.65\,$\pm$\, 0.10 &  
 29.20\,$\pm$\,  1.04 &   2.24$^{+ 0.13}_{- 0.13}$ \\
2846944 &  4326$^{+  98}_{- 145}$&      1.42\,$\pm$\,     0.01& 
      7.9\,$\pm$\,      0.3& 1.79\,$\pm$\,0.02& 0.93\,$\pm$\, 0.07 &  
 20.31\,$\pm$\,  0.85 &   0.46$^{+ 0.01}_{- 0.01}$ \\
2847114 &  4256$^{+  85}_{- 120}$&      1.33\,$\pm$\,     0.02& 
      7.6\,$\pm$\,      0.3& 1.77\,$\pm$\,0.02& 1.03\,$\pm$\, 0.07 &  
 21.95\,$\pm$\,  0.92 &   1.16$^{+ 0.05}_{- 0.05}$ \\
2850913 &  4619$^{+  92}_{- 135}$&      5.23\,$\pm$\,     0.01& 
     49.3\,$\pm$\,      0.7& 2.60\,$\pm$\,0.01& 1.34\,$\pm$\, 0.05 &  
  9.62\,$\pm$\,  0.19 &   1.80$^{+ 0.07}_{- 0.04}$ \\
2853089 &  4972$^{+  99}_{- 171}$&     10.78\,$\pm$\,     0.04& 
    122.1\,$\pm$\,      1.4& 3.01\,$\pm$\,0.01& 1.26\,$\pm$\, 0.04 &  
  5.82\,$\pm$\,  0.11 &   2.36$^{+ 0.12}_{- 0.10}$ \\
2853469 &  3912$^{+  88}_{- 152}$&      0.44\,$\pm$\,     0.01& 
      1.9\,$\pm$\,      0.0& 1.15\,$\pm$\,0.01& 1.19\,$\pm$\, 0.09 &  
 48.17\,$\pm$\,  2.57 &   2.15$^{+ 0.07}_{- 0.07}$ \\
2861062 &  4973$^{+ 168}_{- 145}$&      1.86\,$\pm$\,     0.01& 
     12.9\,$\pm$\,      0.6& 2.03\,$\pm$\,0.02& 1.68\,$\pm$\, 0.15 &  
 20.71\,$\pm$\,  1.06 &   5.35$^{+ 0.34}_{- 0.34}$ \\
2865296 &  4722$^{+  94}_{- 122}$&      3.78\,$\pm$\,     0.14& 
     27.7\,$\pm$\,      0.6& 2.35\,$\pm$\,0.01& 0.90\,$\pm$\, 0.08 &  
 10.48\,$\pm$\,  0.60 &   3.25$^{+ 0.25}_{- 0.19}$ \\
2986860 &  4449$^{+  89}_{- 170}$&      1.49\,$\pm$\,     0.02& 
      9.8\,$\pm$\,      0.7& 1.89\,$\pm$\,0.03& 1.54\,$\pm$\, 0.19 &  
 23.36\,$\pm$\,  1.65 &  14.30$^{+ 1.00}_{- 0.70}$ \\
 ...
 \enddata
\tablenotetext{+}{\footnotesize $T_{\rm eff}*$, \logg, $M_{*}$, $R_*$, and d are respectively the values of the revised effective temperature from \S~3.3, surface gravity, mass, radius from scaling relations and distance from isochrone fitting. }
\label{tbl2}
\end{deluxetable*}%

\begin{deluxetable*}{rcccccr}
\tablewidth{0pt}
\tablecaption{Properties of the red giants with modes near or above the Nyquist frequency from the A2Z+ pipeline\tablenotemark{+}}
\tablehead{\colhead{ KIC} & \colhead{$K_p$} & \colhead{$T_{\rm eff}$ (K) }& \colhead{$\Delta \nu$ ($\mu$Hz)} & \colhead{$\nu_{\rm max}$ ($\mu$Hz)} &  \colhead{$\log g$} & \colhead{Crowding} }
\startdata
1296817 &  12.2& 4931\,$\pm$\,  98&  23.12\,$\pm$\, 0.65& 325.0\,$\pm$\,22.5& 
3.43\,$\pm$\,0.03 &  0.99 \\
3215881 &  12.0& 5045\,$\pm$\, 100&  20.98\,$\pm$\, 0.71& 285.0\,$\pm$\,23.1& 
3.38\,$\pm$\,0.03 &  1.00 \\
3831261 &  14.2& 4921\,$\pm$\,  98&  14.34\,$\pm$\, 0.08& 285.0\,$\pm$\,20.3& 
3.37\,$\pm$\,0.03 &  0.99 \\
3947028 &  15.4& 4750\,$\pm$\,  95&  18.88\,$\pm$\, 0.20& 285.0\,$\pm$\,20.3& 
3.37\,$\pm$\,0.03 &  0.99 \\
4556618 &  14.1& 4911\,$\pm$\,  98&  19.20\,$\pm$\, 0.81& 285.0\,$\pm$\,20.3& 
3.37\,$\pm$\,0.03 &  0.99 \\
4920027 &  15.3& 5180\,$\pm$\, 103&  17.16\,$\pm$\, 0.98& 285.0\,$\pm$\,20.3& 
3.39\,$\pm$\,0.03 &  0.94 \\
5184359 &  14.1& 5158\,$\pm$\, 103&  19.87\,$\pm$\, 0.81& 285.0\,$\pm$\,20.3& 
3.38\,$\pm$\,0.03 &  1.00 \\
5620827 &  11.1& 4998\,$\pm$\,  99&  19.53\,$\pm$\, 2.22& 285.0\,$\pm$\,20.3& 
3.38\,$\pm$\,0.03 &  1.00 \\
5702338 &  14.6& 5078\,$\pm$\, 101&  20.60\,$\pm$\, 0.36& 285.0\,$\pm$\,20.3& 
3.38\,$\pm$\,0.03 &  0.98 \\
6263730 &  11.8& 5151\,$\pm$\, 103&  20.23\,$\pm$\, 0.75& 285.0\,$\pm$\,20.3& 
3.38\,$\pm$\,0.03 &  1.00 \\
6284113 &  11.5& 4975\,$\pm$\,  99&  18.57\,$\pm$\, 0.38& 285.0\,$\pm$\,20.3& 
3.38\,$\pm$\,0.03 &  0.99 \\
6591545 &  14.9& 5132\,$\pm$\, 102&  18.57\,$\pm$\, 0.28& 285.0\,$\pm$\,20.3& 
3.38\,$\pm$\,0.03 &  1.00 \\
7122721 &  13.4& 5029\,$\pm$\, 100&  17.16\,$\pm$\, 0.34& 285.0\,$\pm$\,20.3& 
3.38\,$\pm$\,0.03 &  0.60 \\
7550798 &  13.9& 5075\,$\pm$\, 101&  26.34\,$\pm$\, 0.93& 370.0\,$\pm$\,20.3& 
3.49\,$\pm$\,0.03 &  0.99 \\
7678714 &  13.6& 5093\,$\pm$\, 101&  20.91\,$\pm$\, 0.24& 285.0\,$\pm$\,20.3& 
3.38\,$\pm$\,0.03 &  0.98 \\
7732398 &  13.9& 5385\,$\pm$\, 107&  19.20\,$\pm$\, 0.50& 285.0\,$\pm$\,20.3& 
3.39\,$\pm$\,0.03 &  0.98 \\
8037501 &  13.4& 4840\,$\pm$\,  96&  17.70\,$\pm$\, 0.52& 285.0\,$\pm$\,20.3& 
3.37\,$\pm$\,0.03 &  0.97 \\
8417876 &  14.3& 5228\,$\pm$\, 104&  20.60\,$\pm$\, 0.50& 285.0\,$\pm$\,20.3& 
3.39\,$\pm$\,0.03 &  0.99 \\
8553989 &  12.5& 5132\,$\pm$\, 102&  24.10\,$\pm$\, 0.55& 285.0\,$\pm$\,20.3& 
3.38\,$\pm$\,0.03 &  0.99 \\
9095475 &  14.7& 5035\,$\pm$\, 100&  21.37\,$\pm$\, 0.74& 285.0\,$\pm$\,20.3& 
3.38\,$\pm$\,0.03 &  0.98 \\
9221660 &  12.4& 5160\,$\pm$\, 103&  19.87\,$\pm$\, 0.49& 285.0\,$\pm$\,20.3& 
3.38\,$\pm$\,0.03 &  1.00 \\
9637300 &  13.6& 5095\,$\pm$\, 101&  23.60\,$\pm$\, 0.96& 335.0\,$\pm$\,20.3& 
3.45\,$\pm$\,0.03 &  1.00 \\
9639069 &  14.4& 5193\,$\pm$\, 103&  19.53\,$\pm$\, 0.30& 285.0\,$\pm$\,20.3& 
3.39\,$\pm$\,0.03 &  1.00 \\
9698043 &  13.9& 5178\,$\pm$\, 103&  22.66\,$\pm$\, 0.41& 285.0\,$\pm$\,20.3& 
3.39\,$\pm$\,0.03 &  0.99 \\
9706054 &  14.3& 5205\,$\pm$\, 104&  21.78\,$\pm$\, 0.53& 285.0\,$\pm$\,20.3& 
3.39\,$\pm$\,0.03 &  0.97 \\
10264711 &  14.3& 5078\,$\pm$\, 101&  20.60\,$\pm$\, 0.49& 285.0\,$\pm$\,20.3& 
3.38\,$\pm$\,0.03 &  0.99 \\
11138117 &  14.6& 5159\,$\pm$\, 103&  22.21\,$\pm$\, 0.34& 285.0\,$\pm$\,20.3& 
3.38\,$\pm$\,0.03 &  1.00 \\
11181529 &  13.3& 5209\,$\pm$\, 104&  20.23\,$\pm$\, 0.48& 285.0\,$\pm$\,20.3& 
3.39\,$\pm$\,0.03 &  1.00 \\
11250788 &  11.1& 5375\,$\pm$\, 107&  19.87\,$\pm$\, 0.64& 285.0\,$\pm$\,20.3& 
3.39\,$\pm$\,0.03 &  1.00 \\
11443679 &  13.7& 5361\,$\pm$\, 107&  19.20\,$\pm$\, 0.39& 285.0\,$\pm$\,20.3& 
3.39\,$\pm$\,0.03 &  0.96 \\
11666333 &  13.8& 5254\,$\pm$\, 105&  19.20\,$\pm$\, 0.46& 285.0\,$\pm$\,25.0& 
3.39\,$\pm$\,0.03 &  0.96 
\enddata
\tablenotetext{+}{\footnotesize $T_{\rm eff}$ and $\logg$ are respectively the values of the effective temperature from the KIC and surface gravity from scaling relations.}
\label{tbl3}
\end{deluxetable*}%

\subsection{Revised Effective Temperatures and Distances}

\begin{figure*}[htbp]
\begin{center}
\includegraphics[angle=90,width=14cm]{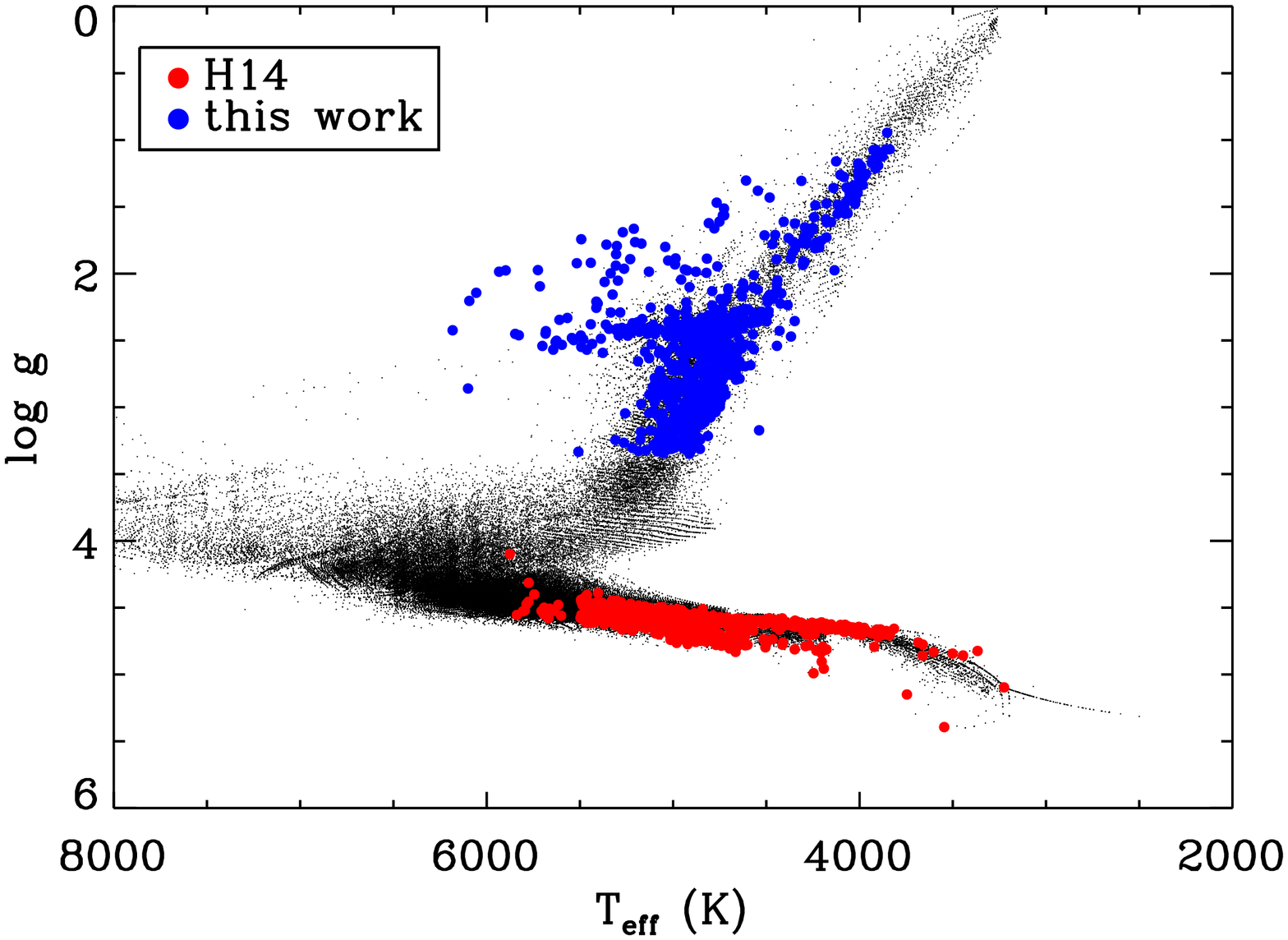}
\caption{HR diagram of the stars cataloged in H14 (black dots). Stars in our sample are overplotted with colored circles: the old stellar parameters are represented in red symbols and the parameters derived in this work (i.e. seismic $\logg$ and revised $T_{\rm eff}$) are represented with blue symbols.}
\label{Fig_new_Teff}
\end{center}
\end{figure*}

Previous estimates of effective temperatures for this sample were mostly 
based on matching broadband photometry to models assuming a simple 
reddening law as part of the classifications for the KIC. If the stars are indeed giants instead of dwarfs, 
these temperatures will be underestimated since the stars are 
more distant and therefore more reddened than initially assumed.

To determine revised temperatures and distances for the sample 
we used broadband photometry, asteroseismic observables, and a grid of Parsec 
isochrones \citep{2012MNRAS.427..127B}. The grid ranges 
from 0.5--14\,Gyr with a stepsize of 0.5\,Gyr and $Z=0.0001-0.03$ with a stepsize of 
0.0001 (corresponding to metallicities between $\feh=-2.18$ and $\feh=+0.42$), 
and was calculated with a Reimers mass loss parameters set to $\eta=0.2$. 

Given a set of observables  $x=\{J-K,H-K,g-r,r-i,\numax,\dnu\}$
with Gaussian uncertainties $\sigma_x$ and a set of 
intrinsic parameters $y=\{\rm age,\ [Fe/H],\ mass\}$, we calculated the 
posterior probability of the observed star
having intrinsic parameters $y$ as:

\begin{equation}
\begin{split}
p(y|x) \propto p(y)p(x|y) \propto  p(y)\prod_i \exp{\left(-\frac{(x_i-x_i(y))^2}{2 \sigma_{x,i}^2}\right)} \: .
\end{split}
\end{equation}

For $p(y)$ we adopt a flat prior on age and mass, and a metallicity prior derived from
the Geneva Copenhagen survey \citep{2011A&A...530A.138C}, 
which has been shown to agree well with the metallicity distribution of {\it Kepler} 
targets \citep{2014ApJ...789L...3D}. We used 2MASS $J-H$ 
and $H-K$ colors if all three bands have the highest photometric quality flag 
(Qflg='AAA'), and KIC $g-r, r-i$ colors otherwise. For the KIC photometry we interpolated 
uncertainties as a function of $Kp$ using typical values taken from Figure~\ref{Fig_new_Teff} 
in \citet{2011AJ....142..112B}. {  Nine} stars did not have all the colors available or a not high-enough quality flag so they do not have results from this method. Model colors and distances are calculated 
from absolute magnitudes, which are reddened using the 3D extinction map by 
\citet{2005AJ....130..659A} for 
distances corresponding to the apparent magnitude (either $J$-band or $g$-band) and 
galactic coordinates of a 
given target. Values for \numax\ and \dnu\ values for each model were 
calculated using the scaling relations \citep{1995A&A...293...87K}, 
assuming solar reference values of $\numax=3090\muHz$ and $\dnu=135.1\muHz$ \citep{2011ApJ...743..143H}.  
Temperatures and distances were then derived from the median and $1\sigma$ 
confidence interval of the probability distribution function obtained by integrating 
$p(y|x)$ along the isochrone grid, weighted by local grid spacing in 
mass, age and metallicity \citep{2013MNRAS.429.3645S}. 
We note that we assumed that the targets were single stars when deriving the distances but that effect should be small.

For the stars with $\nu_{\rm max}$ below the Nyquist frequency, we computed the mass, radius, and $\log g$ using the scaling relations along with the updated effective temperature. The stellar parameters are all listed in Tables~\ref{tbl1} and \ref{tbl2} , respectively for the A2Z+ and the SYD pipelines. For the stars with modes near or above the Nyquist frequency, we only computed the $\log g$ as we only have a very rough estimate of $\numax$. The seismic parameters for this subsample of stars are given in Table~\ref{tbl3}.
Figure~\ref{Fig_new_Teff} shows a $\logg$ versus $\teff$ diagram comparing the positions of the 
sample before and after the re-derivation of $\logg$ and $\teff$. Note that $\logg$ values for 
the sample were calculated using the $\numax$ scaling relations with the revised 
$\teff$. As expected, most stars shift to hotter temperatures since the de-reddened 
colors assuming larger distances are bluer than de-reddened colors assuming that the 
stars are dwarfs.

\subsection{Mixed modes and evolutionary stage}



\begin{figure}[htbp]
\begin{center}
\includegraphics[width=9cm]{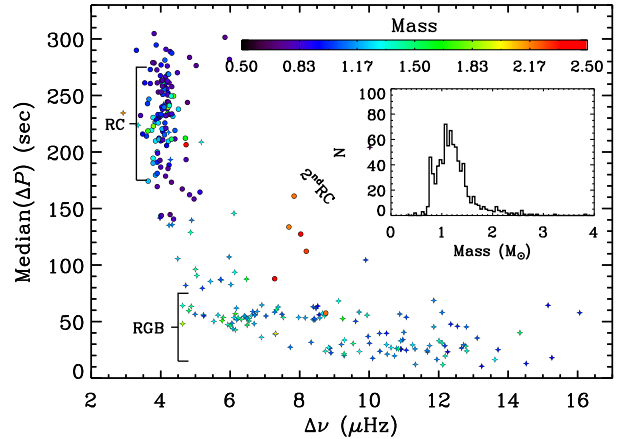}
\caption{Median period spacing versus $\Delta\nu$ for the 280 stars providing a
clear classification into either red giant branch or red clump (including
secondary clump). Dots show the helium core-burning stars, and stars show
the red giant branch stars. The color of the symbols indicate the stellar
mass derived from the $\Delta\nu$ and $\nu_{\mathrm{max}}$ scaling
relations. The mass distribution of the plotted sample is shown in the
inset. }
\label{dp-dnu}
\end{center}
\end{figure}

As stars evolve to become red giants the acoustic modes in the stellar
envelope start to couple with the gravity modes in the core, which leads to
the so-called mixed modes \citep{2009A&A...506...57D}. Mixed modes have been observed in
several thousands of giants in the {\it Kepler} field \citep[e.g.][]{2011Sci...332..205B,2011Natur.471..608B,2012A&A...540A.143M,2013ApJ...765L..41S}.
Because of their partly gravity-mode
nature, they are sensitive to the stellar core properties, and in
particular whether a star is ascending the red giant branch (with an
inert core and only burning hydrogen in a shell) or if it is a red clump
star also burning helium in the core \citep{2011Natur.471..608B,2014A&A...572L...5M}.
This classification into red giant branch and red clump stars shows up as a
difference in the period spacing of consecutive overtone mixed modes.
We therefore attempted to measure the median period spacing for each star
in our sample using the method by \citet{2013ApJ...765L..41S}. Figure~\ref{dp-dnu}
shows the median period spacing versus $\Delta\nu$ for the 280 stars where
we could obtain a clear classification after verification by visual
inspection of the power spectra. Some of the red giant branch stars at low
$\Delta\nu$ are artificially pushed up to larger period spacings by the
method.  In this region between the red clump and the red giant branch the
classifications, shown by symbol type, relies on the visual inspection of the
power spectra. Compared to the large `typical' {\it Kepler} sample of giants
analyzed by \citet[][their Fig.~4a]{2013ApJ...765L..41S}, we see that our new sample
here is strongly dominated by low-mass stars; only very few clump stars with
masses above about 2M$_\odot$ are present (marked 2ndRC). 
The classifications along with the the value of the median period
spacings are given in Table 4.  We note 
that given the method to determine the median period spacings they are not
suitable for direct model comparison, and serve here merely as a way to
classify the stellar evolutionary state.

\begin{deluxetable}{rcc}
\tablewidth{0pt}
\tablecaption{Evolutionary stage and period spacing for 280 misclassified red giants}
\tablehead{\colhead{ KIC} &  \colhead{Median $\Delta P$ (s)} & \colhead{Evolutionary Stage}  }
\startdata
1160684 &  214.0 &RC/2ndRC \\
1431059 &   16.3 &RGB \\
2010577 &  190.9 &RC/2ndRC \\
2014737 &  230.8 &RC/2ndRC \\
2159728 &   60.4 &RGB \\
2297384 &  230.5 &RC/2ndRC \\
2297825 &  190.1 &RC/2ndRC \\
2308968 &  276.4 &RC/2ndRC \\
2436209 &   86.7 &RGB \\
2436540 &   53.7 &RGB \\
2437402 &   59.6 &RGB \\
2437539 &   63.4 &RGB \\
2437589 &   48.1 &RGB \\
2437698 &  219.3 &RC/2ndRC \\
2437805 &  179.6 &RC/2ndRC \\
2438038 &   42.7 &RGB \\
2570094 &   46.9 &RGB \\
2708445 &   53.5 &RGB \\
2850913 &   56.0 &RGB \\
2865296 &  230.6 &RC/2ndRC \\
3110892 &  260.0 &RC/2ndRC \\
3239099 &  255.8 &RC/2ndRC \\
3331116 &  231.7 &RC/2ndRC \\
3432568 &   43.7 &RGB \\
3534438 &  283.7 &RC/2ndRC \\
3543446 &   23.2 &RGB \\
3952678 &  238.2 &RC/2ndRC \\
3953036 &   64.6 &RGB \\
3956618 &  143.4 &RC/2ndRC \\
3957962 &  244.4 &RC/2ndRC \\
3964353 &   54.2 &RGB \\
3965974 &  262.8 &RC/2ndRC \\
3972205 &   47.3 &RGB \\
3972697 &  266.5 &RC/2ndRC \\
3974206 &  226.5 &RC/2ndRC \\
4041075 &  194.5 &RC/2ndRC \\
4041406 &  112.2 &RC/2ndRC \\
4041741 &   50.5 &RGB \\
4049147 &   90.7 &RGB \\
4057657 &  257.8 &RC/2ndRC \\
4067867 &  249.5 &RC/2ndRC \\
4141587 &  200.1 &RC/2ndRC \\
4141698 &   54.4 &RGB \\
4246729 &   64.1 &RGB \\
4248675 &  169.4 &RC/2ndRC \\
4254250 &  239.8 &RC/2ndRC \\
4254422 &  212.2 &RC/2ndRC \\
4265786 &   12.1 &RGB \\
4271799 &  278.8 &RC/2ndRC \\
4366888 &   31.1 &RGB \\
4446637 &   12.9 &RGB \\
...
\enddata
\label{tbl4}
\end{deluxetable}%


\section{Discussion of the sample}


\subsection{Faint stars and distances}

\begin{figure}[htbp]
\begin{center}
\includegraphics[angle=90,width=9cm]{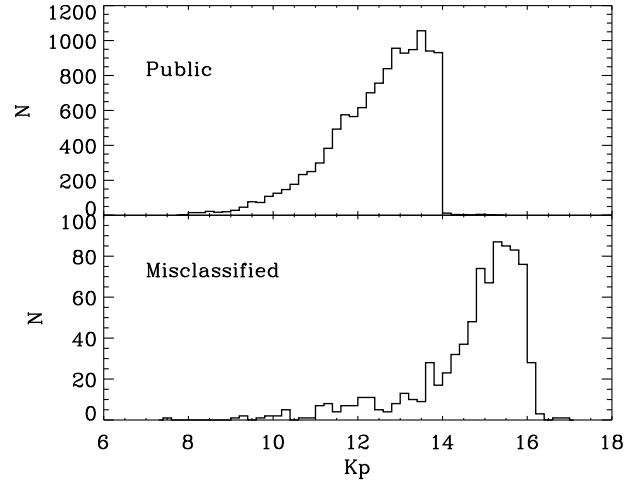}
\caption{Comparison of the magnitude distribution of our confirmed red giants (bottom panel) and the public red giants (top panel).}
\label{Fig_histo_mag}
\end{center}
\end{figure}

We looked at the distribution of some of the stellar properties of the misclassified red giant sample. Figure~\ref{Fig_histo_mag} shows the distribution of the {\it Kepler} magnitude for {  the confirmed} red giants and for the $\sim$\, 13,000 public red giants in \citet{2011ApJ...743..143H}. We clearly see that the new red giant sample is very different from the known red giants one: they are much fainter than the previously known sample. 

Even though the stars are so faint, we are still able to detect solar-like oscillations and characterize their global properties. It seems from Figure~\ref{Fig_histo_mag} that {\it Kepler} would have been able to detect oscillations in giants even fainter than $Kp\,=\,16$ if they had been observed. This result is very promising for missions like K2 \citep{2014PASP..126..398H} and its galactic archeology program \citep{2015ApJ...809L...3S} and  Transiting Exoplanet Survey Satellite \citep[TESS][]{2015JATIS...1a4003R}.



\begin{figure}[htbp]
\begin{center}
\includegraphics[angle=90,width=8cm]{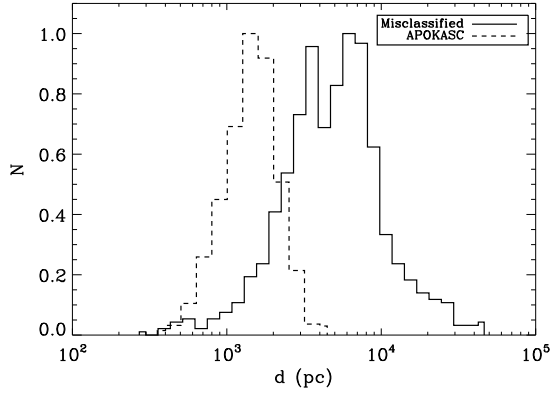}
\caption{Normalized distribution of {  the distances} in parsec for the confirmed red giants of our sample (solid line) compared to the distances of the APOKASC red giants (dashed line) from \citet{2014MNRAS.445.2758R}.}
\label{histo_dist}
\end{center}
\end{figure}

\begin{figure}[htbp]
\begin{center}
\includegraphics[width=9cm]{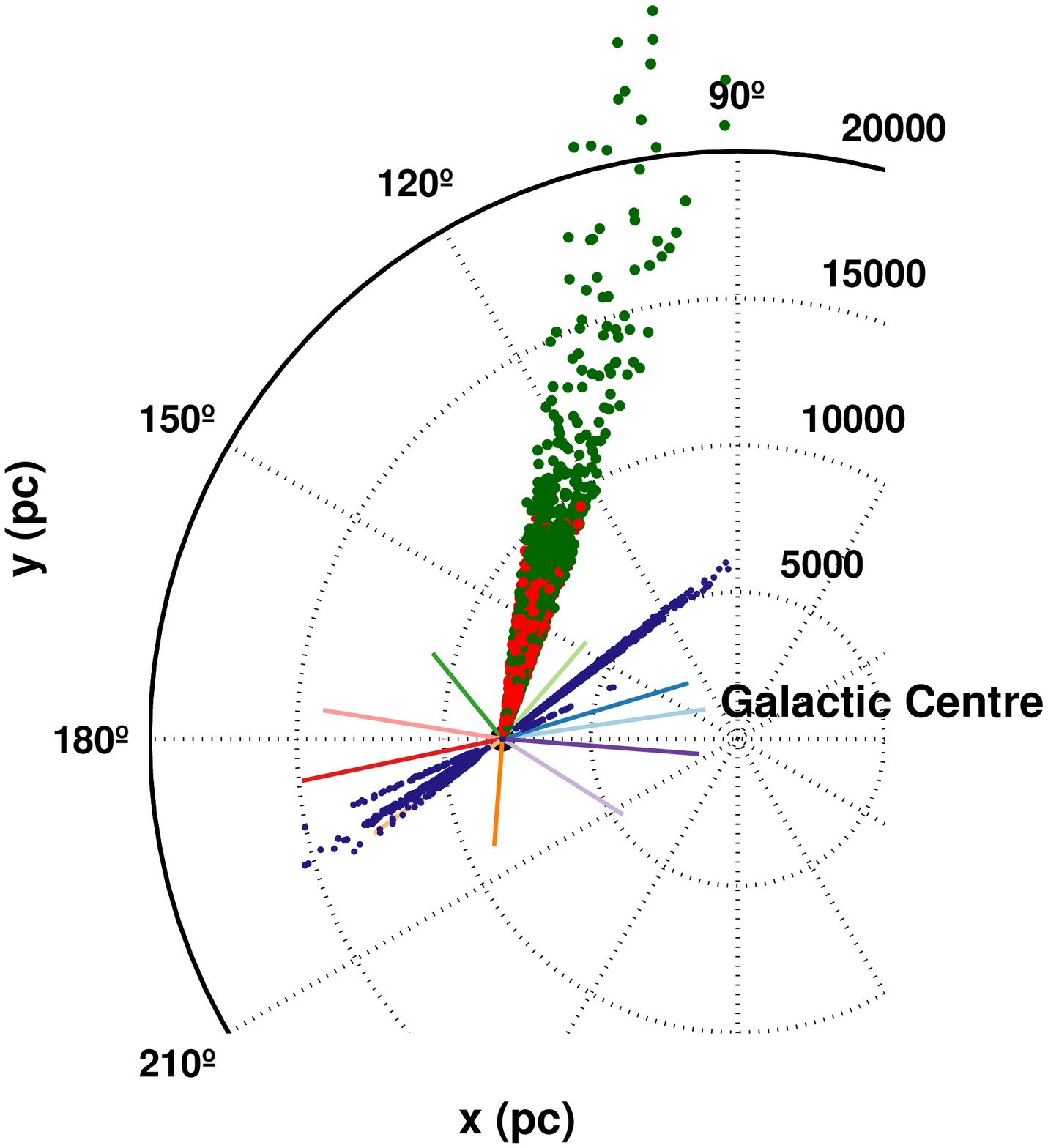}
\includegraphics[width=9cm]{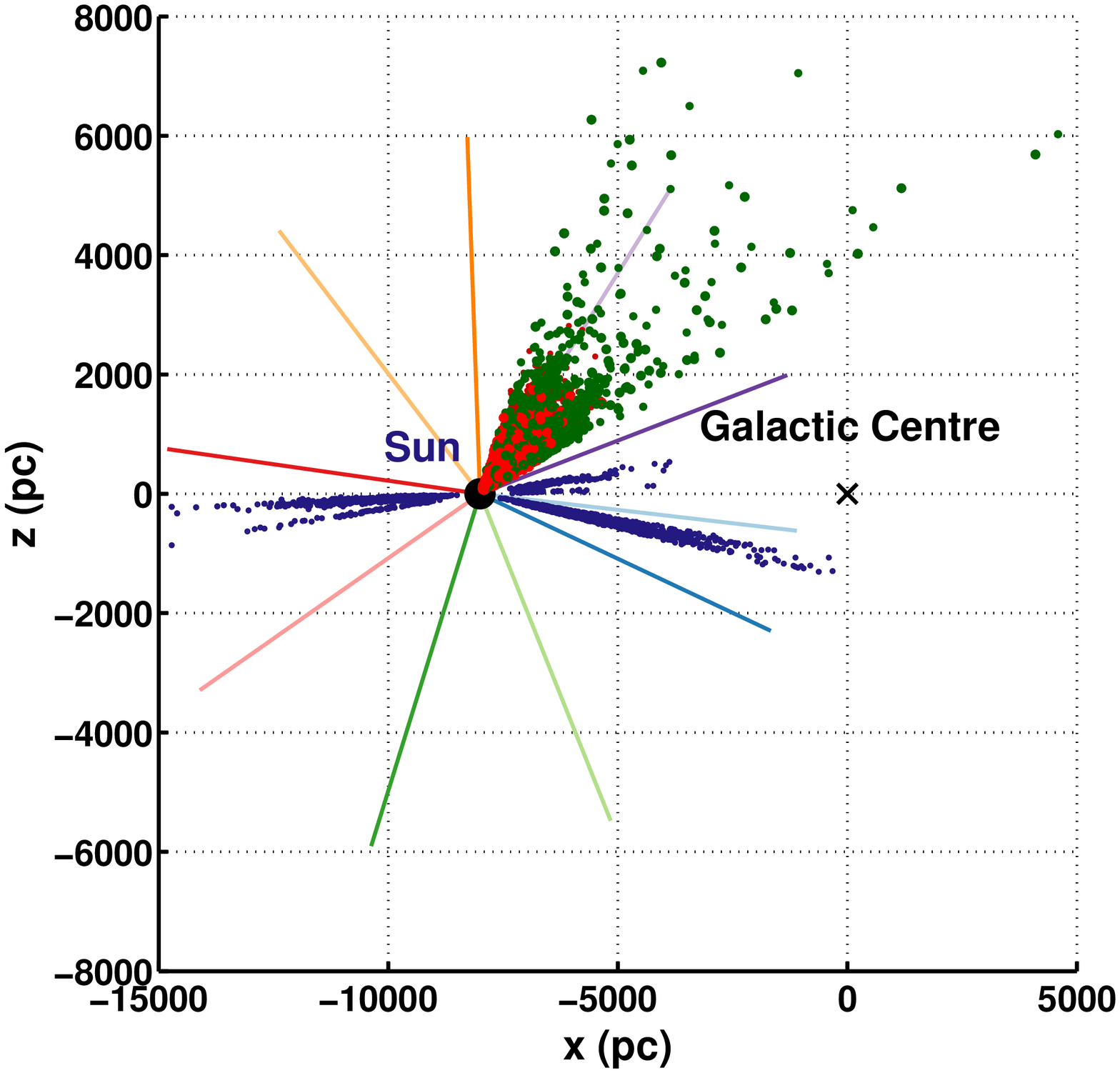}
\includegraphics[width=9cm, trim=0 4cm 0 0]{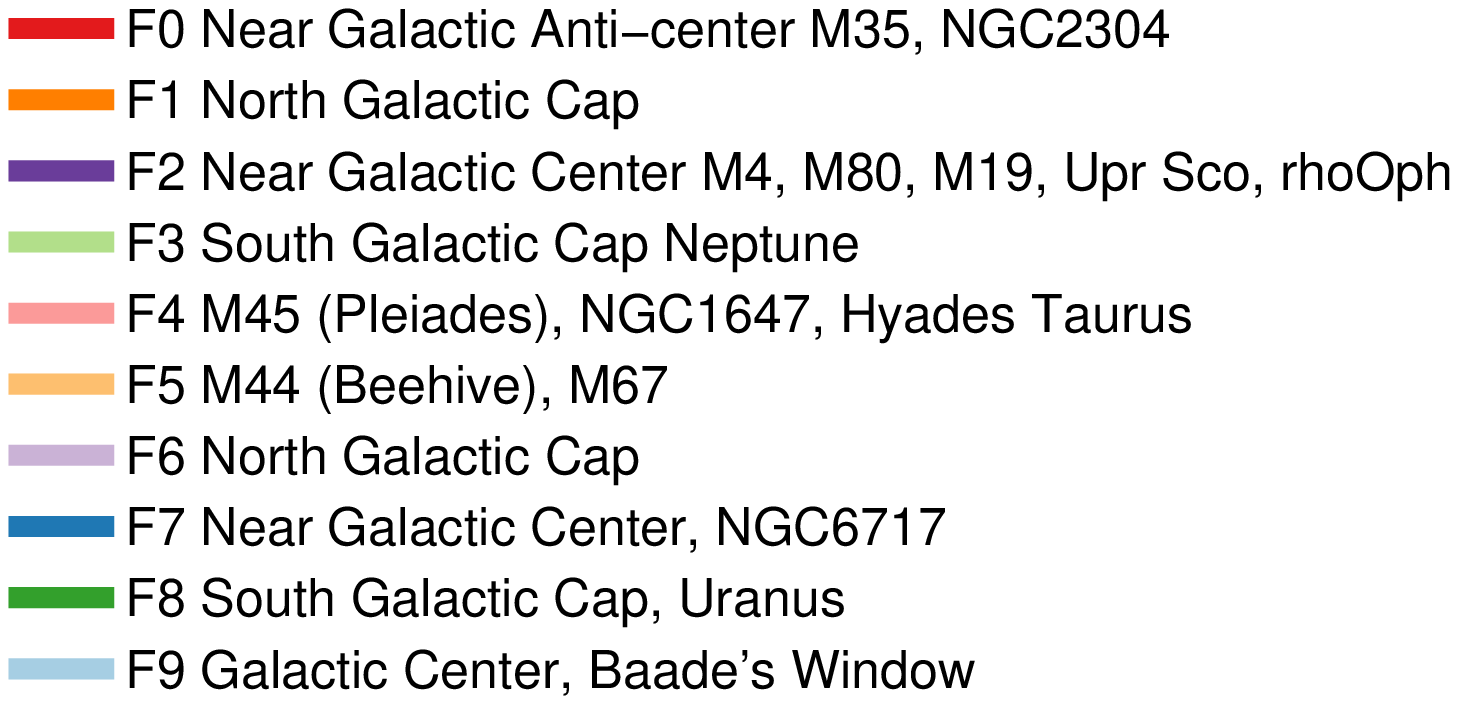}
\caption{Position in the Galaxy of the new red giants (green circles) compared to the stars in the CoRoT (blue dots), {\it Kepler} (red dots), and K2 fields (solid lines) {  as a function of x and y (top panel) and x and z (bottom panel)}.}
\label{position_diag}
\end{center}
\end{figure}

Figure~\ref{histo_dist} shows the distance distribution of the newly classified red giants compared to another sample of red giants observed by {\it Kepler}. 
\citet{2014MNRAS.445.2758R} determined the distances of $\sim$~\,2000 {\it Kepler} red giants that also had spectroscopic observations with the Apache Point Observatory Galactic Evolution Experiment \citep[APOGEE][]{2015arXiv150905420M}. Their sample of stars (hereafter called the APOKASC sample) had distances up to $\sim$\,4\,kpc.
We note that the distances of the new red giants are quite large compared to the APOKASC sample and go well beyond 10~kpc. Our sample of new red giants has 404 stars at a distance larger than 5~kpc. This new sample allows us to probe our Galaxy further away and study stellar populations with seismology in a different region compared to what has been done so far with the CoRoT and {\it Kepler} missions. This can be seen in Figure~\ref{position_diag}, which is an update of the Figure 2 of \citet{2015ASSP...39...11M}.

\begin{figure}[htbp]
\begin{center}
\includegraphics[angle=90,width=8cm]{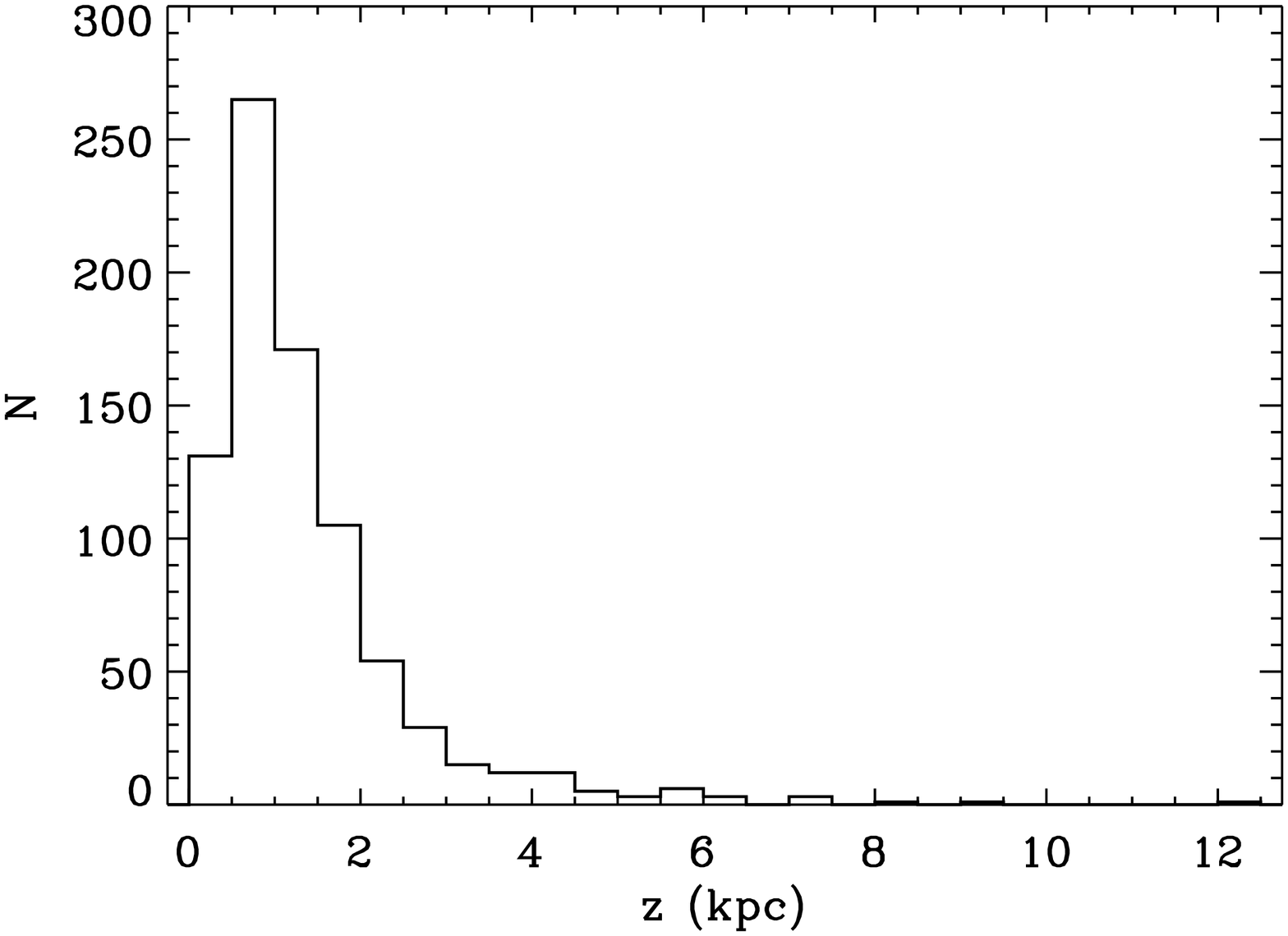}
\caption{{  Distribution of the heights of the misclassified red giants.}}
\label{histo_height}
\end{center}
\end{figure}

{  We also computed the heights of the stars with respect to the Galactic plane to infer the possible number of halo stars present in our sample. Their position in the Galaxy are shown in the bottom panel of Figure~\ref{position_diag}. Figure~\ref{histo_height} shows the distribution of the heights of the misclassified stars. We compared this distribution with a synthetic population of halo stars in the Kepler field with $J<16$ generated using Galaxia  \citep{2011ApJ...730....3S}, which shows that the majority of stars ($\gtrsim$60\%) above 5 kpc are halo stars, while the faction of halo stars drops rapidly towards the galactic plane ($\lesssim$2\% below 2 kpc). We divided the fraction of halo stars from the Galaxia model and our observed sample in height bins of 2 kpc to estimate the number of potential halo stars in each bin, yielding a total of about 40 red giants in our sample which could be halo stars. We note that this estimate is statistical only, and may be affected by systematic errors in the Galaxy model.}





\subsection{Mass distribution of the sample}

\begin{figure}[htbp]
\begin{center}
\includegraphics[angle=90,width=8cm]{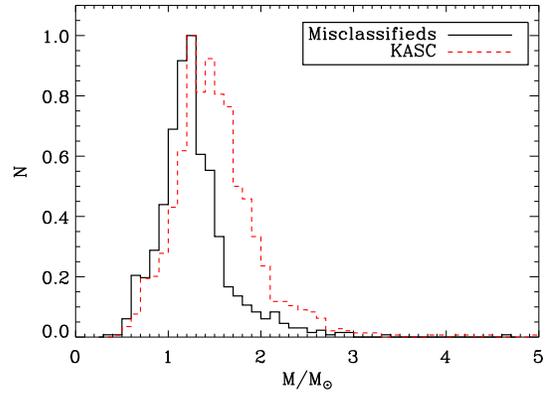}
\includegraphics[angle=90,width=8cm]{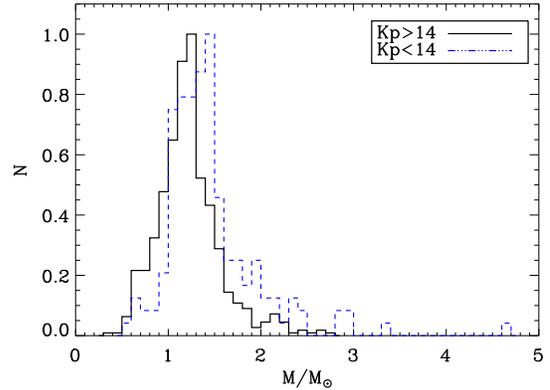}
\includegraphics[angle=90,width=8cm]{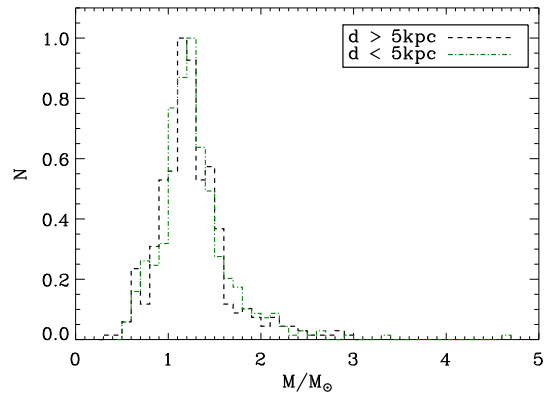}
\caption{Distribution of mass for the misclassified red giants. Top panel: mass histograms of our sample (black solid line) compared to the KASC sample (red dashed line). Middle panel: mass histograms for stars brighter (blue dotted-cached line) or fainter (black solid line) than a {\it Kepler} magnitude of 14. Bottom panel: mass histograms for stars in our sample which are closer than 5 kpc (red dashed-dotted line) and beyond 5 kpc (black dashed line).}
\label{histo_mass}
\end{center}
\end{figure}

Figure~\ref{histo_mass} top panel presents the mass distribution of the newly classified red giants and the KASC sample of red giants \citep[e.g.][]{2011ApJ...743..143H} (red dashed line). Both samples peak at around 1.2\,$M_{\odot}$. However, we clearly see that the number of high-mass (M larger than 1.3$M_{\odot}$) stars is smaller in our new sample compared to the KASC one as already noted in Section 3.4. There are slightly more stars with masses below 1.1\,$M_\odot$.  This agrees with the fact that these stars are more distant, and thus possibly halo stars. In the middle panel of Figure~\ref{histo_mass} we also note that the low-mass stars are in general fainter ($K_p>$14).

In fact, the generally lower mass of the sample would imply larger expected amplitudes at a given $\numax$ \citep{2011ApJ...737L..10S}. In Figure~\ref{Amax_numax}, we note that there are some stars with slightly higher mode amplitude ($A_{\rm max}$) compared to the general trend. We looked at the masses of the low-amplitude mode stars but did not find any systematic low-mass stars. 

{  Given that some of these stars could belong to the halo and are thus metal-poor stars, some of the masses derived with the scaling relations should be taken cautiously. Indeed \citet{2014ApJ...785L..28E} studied a sample of halo stars observed by {\it Kepler} and showed that the scaling relations seem to deviate. As we do not have reliable metallicity measurement for these stars yet, we cannot take out the metal-poor stars from the sample.}

{  We note that there are two stars (KIC 4850755 and 6266309) with a mass below 0.4M$_\odot$. Their parameters seem correct although KIC 4850755 seems a little odd from the visual inspection with a low number of modes observed and low amplitude for the l=1 modes as reported for example by \citet{2014A&A...563A..84G} and theoretically explained by \citet{2015Sci...350..423F} and \citet{2016Natur.529..364S}. It could also be that scaling relations are not valid for such low-mass stars. More stars of this type would be needed to verify the validity of the scaling relations in this mass regime.}

We also looked at differences in the mass distribution as a function of distance (see bottom panel of Figure~\ref{histo_mass}). We divided the sample into stars with distances larger than 5\,kpc and stars with distances smaller than 5\,kpc. The general distribution of both samples is very similar.


\subsection{Possible explanations for dwarf classification}

We checked the type of data and analysis that were used in the {\it Kepler} H14 catalog, to derive the effective temperature and the surface gravity of the {  854} confirmed red giants. The majority of the stars of our sample have values obtained from the KIC. However, 24 stars have their $\log g$ determined by \citet{2014ApJS..211....2H} based on photometric observations and 8 stars were characterized by the work of \citet{2013ApJ...767...95D} who analyzed optical and near-infrared photometry from the KIC to improve the stellar parameters of the cool {\it Kepler} dwarfs. 

Recently, \citet{2016MNRAS.457.2877G} classified the cool {\it Kepler} dwarfs. We found six stars that overlap with our sample. They have the flag 'PM' for 'Possible M dwarfs'. Four of them either a low crowding value or are possible pollution. 



We aforementioned that the misclassified stars have their parameters determined from photometric observations. Distinguishing the luminosity class between dwarf and red giants can be a difficult task when only using multi-color photometry and the respective color indices. In approximately 4 \% of KIC targets, there were significant discrepancies between different photometric temperature indicators \citep{2012ApJS..199...30P}. This fraction was comparable to the 3.8 \% of targets with significant blends, estimated from a comparison of KIC photometry to that of the higher angular resolution SDSS photometry.  In an even smaller minority of cases, the gravity classification was also in significant error, most likely due to a similar cause. Thus it is not surprising that {  854} stars (less than 0.5\% of the {\it Kepler} targets) were misclassified. This number of stars is actually remarkably low given the large number of targets.

\subsection{Planet-Candidate Host Stars}


Five of the newly classified red giants are planet-candidate hosts stars. In Table~\ref{tbl5}, we compare the new seismic stellar radius with the H14 value. Most of these stars were classified as M-dwarfs but now that they are classified as red giants, their new radii are 7 to 21 times larger than the ones listed in the H14 catalog.
 

We combined our revised stellar radii with the transit depths listed at the NASA Exoplanet 
Archive\footnote{\url{http://exoplanetarchive.ipac.caltech.edu}} to re-calculate the 
planet-candidate radii, also given in Table~\ref{tbl5}. To calculate the uncertainties on the planet radius, we used a symmetric uncertainty on the ratio $R_p/R_*$ where we decided to take the largest values of the positive and negative values and then propagated the error bars. Given the new classification of the host stars, the estimated planet sizes have changed from Earth-size/superEarth-size planets to {  objects bigger than 5$\times R_{+}$}. 


To re-evaluate the planet-candidate 
status we used the revised stellar mass and radius to estimate the expected transit 
duration for a circular orbit and compared this to the observed transit durations. 
For three planet-candidate hosts (KOI-5652, KOI-5859, KOI-2813) the observed durations 
are significantly shorter than expected for a planet orbiting an evolved star, indicating 
that the planet-candidate would have to be either on a highly eccentric orbit or have an 
extremely high impact parameter. Inspection of the data validation (DV) reports showed that 
KOI-5652 and KOI-5859 both have low S/N, and hence are likely spurious 
detections due to the high correlated noise in giants due to stellar 
granulation. KOI-2813 is flagged as a likely false positive according to 
spectroscopic follow-up observations by the {\it Kepler} Community 
Follow-Up\footnote{\url{https://cfop.ipac.caltech.edu/home/}}, which is supported 
by our reclassification of the host star. Finally, KOI-4902.01 and KOI-6217.01 show 
transit durations which are compatible with a low-luminosity RGB star, and hence 
remain viable planet-candidates.

\begin{deluxetable*}{rccccccc}
\tablewidth{0pt}
\tablecaption{Determination of the new planet radius for the planet-candidate host stars reclassified as red giants}
\tablehead{\colhead{ KIC} & \colhead{KOI name} &  \colhead{$R_{\rm p}/R_{\rm *}$ } & \colhead{$R_*$ (new)}&  \colhead{$R_*$ (old)}&  \colhead{$R_p$ (new) }&  \colhead{$R_p$ (old) } &  \colhead{Planet-candidate}\tablenotemark{+} \\
 & & & ($R_\odot$)& ($R_\odot$)& ($R_{+}$)& ($R_{+}$) &  status}
\startdata

 9292100 & K05652  & 0.021 $\pm$ 0.002  & 6.00 $\pm$ 0.44  & 0.63 $\pm$ 2.36  &13.70 $\pm$ 2.08  & 1.43 $\pm$ 5.41& Likely FP\\
10000162 & K04902  & 0.020 $\pm$ 0.003  & 5.44 $\pm$ 0.38  & 0.80 $\pm$ 0.26  &12.10 $\pm$ 2.40  & 1.78 $\pm$ 0.57& PC\\
10200627 & K06217  & 0.011 $\pm$ 0.001  & 4.23 $\pm$ 0.56  & 0.54 $\pm$ 0.04  & 5.01 $\pm$ 0.99  & 0.64 $\pm$ 0.04& PC\\
11100170 & K05869  & 0.032 $\pm$ 0.002  &11.90 $\pm$ 0.89  & 0.69 $\pm$ 0.07  &41.65 $\pm$ 6.06  & 2.42 $\pm$ 0.23& Likely FP\\
11197853 & K02813  & 0.011 $\pm$ 0.001  &10.47 $\pm$ 0.75  & 0.62 $\pm$ 1.88  &12.15 $\pm$ 1.75  & 0.73 $\pm$ 2.18& Likely FP
\enddata
\label{tbl5}
\tablenotetext{+}{\footnotesize FP and PC stand respectively for 'False-Positive' and 'Planet Candidate'.}
\end{deluxetable*}%

\section{Conclusions}

The analysis of 45,431 dwarfs from the Q1-Q16 {\it Kepler} star properties catalog \citep{2014ApJS..211....2H} led to the re-classification of {  854} of them as red giants, among which 31 stars have modes close to or above the Nyquist frequency. The comparison of the power spectra with known red giants and the inspection of UKIRT J-band images suggest that some of these newly classified red giants could still result from the presence of a nearby star. 

We analyzed the light curves with two different pipelines, A2Z+ and SYD, and measured the global parameters of the acoustic modes. We then computed stellar fundamental parameters such as mass and radius from scaling relations and revised effective temperatures and distances from isochrone fitting for stars {  with modes below Nyquist frequency and for which colors were available}. Among that sample of stars, we determined the evolutionary stage of 280 stars based on the period spacing of mixed modes. 

We find that this sample of new red giants are less massive, fainter, and more distant than the previously known sample of red giants observed by {\it Kepler}, {  extending the} parameter space. This work demonstrates that we can push the mode-detection limits in red giants to much fainter stars up to $Kp$ around 16.

These faint red giants represent a goldmine for galactic archeology studies \citep[e.g.][]{2013MNRAS.429..423M,2015ApJ...809L...3S} as these stars probe the edge of the Milky Way. They will be observed by APOGEE in order to study their composition and have additional information to better understand the chemical evolution of the Galaxy.

\acknowledgments
The authors would like to thank A. Miglio for his help to make Figure~\ref{position_diag} of this paper. The authors would like to thank M. H. Pinsonneault for useful discussions. The authors are also thankful to J. L. van Saders for her help with the UKIRT J-band images. SM acknowledges support from the NASA grant NNX12AE17G. SM and DH acknowledge support by the National Aeronautics 
and Space Administration under Grant NNX14AB92G issued through the Kepler Participating 
Scientist Program. RAG, PGB, and KH acknowledge the support of the European Community's Seventh Framework Programme (FP7/2007-2013) under grant agreement no. 269194 (IRSES/ASK). RAG and PGB acknowledge the support by the French ANR/IDEE grant. This research was supported in part by the National Science Foundation under Grant No. NSF PHY11-25915. D.H. acknowledges support by the Australian Research Council's Discovery Projects 
funding scheme (project number DE140101364). This work has received funding from the CNES.

\bibliographystyle{apj} 

\bibliography{apj-jour,/Users/Savita/Documents/BIBLIO_sav}


\appendix

\section{Comparison of the A2Z+ and SYD pipelines}
\subsection{Seismic parameters}

\begin{figure}[http]
\begin{center}
\includegraphics[angle=90,width=8.5cm]{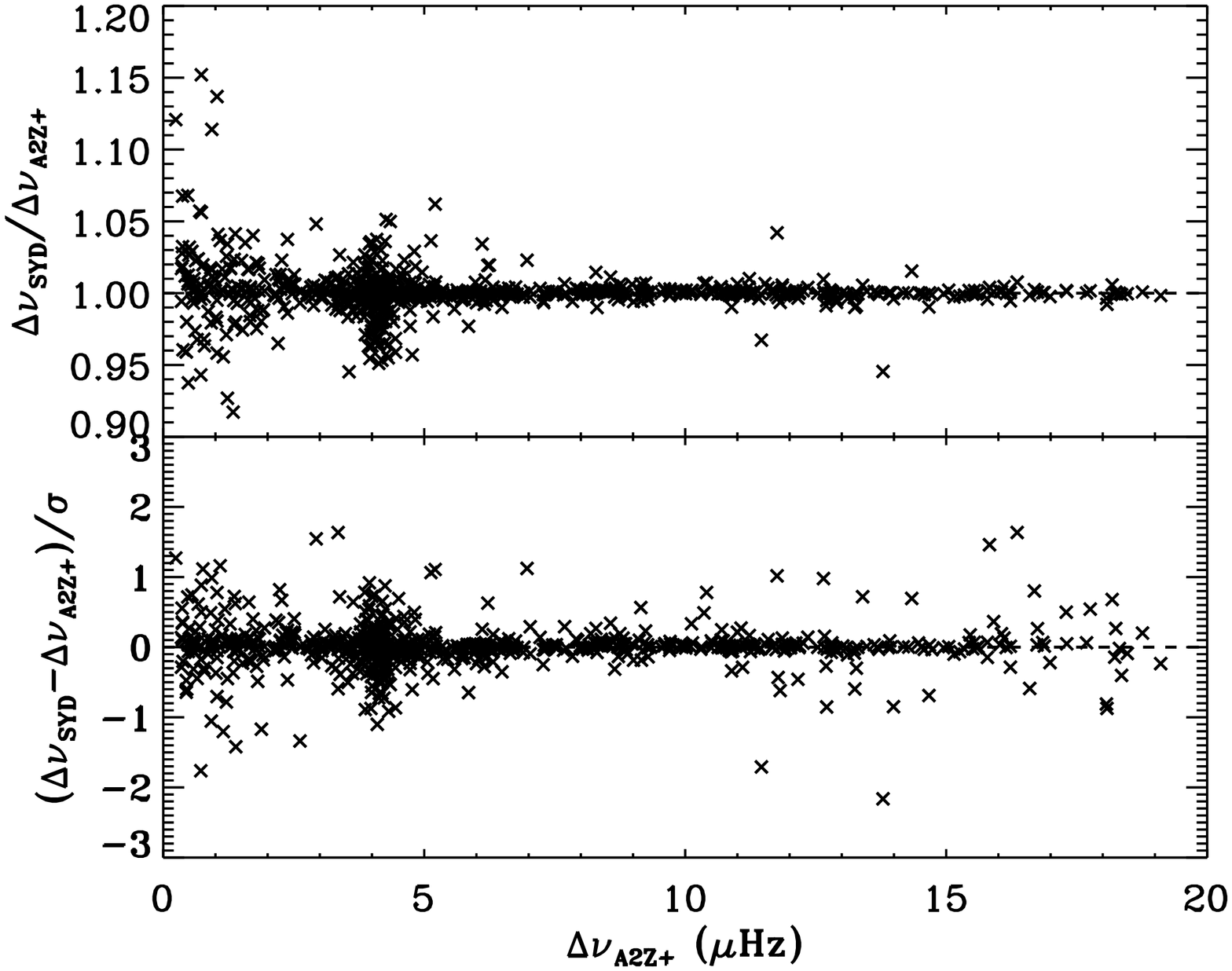}
\includegraphics[angle=90,width=8.5cm]{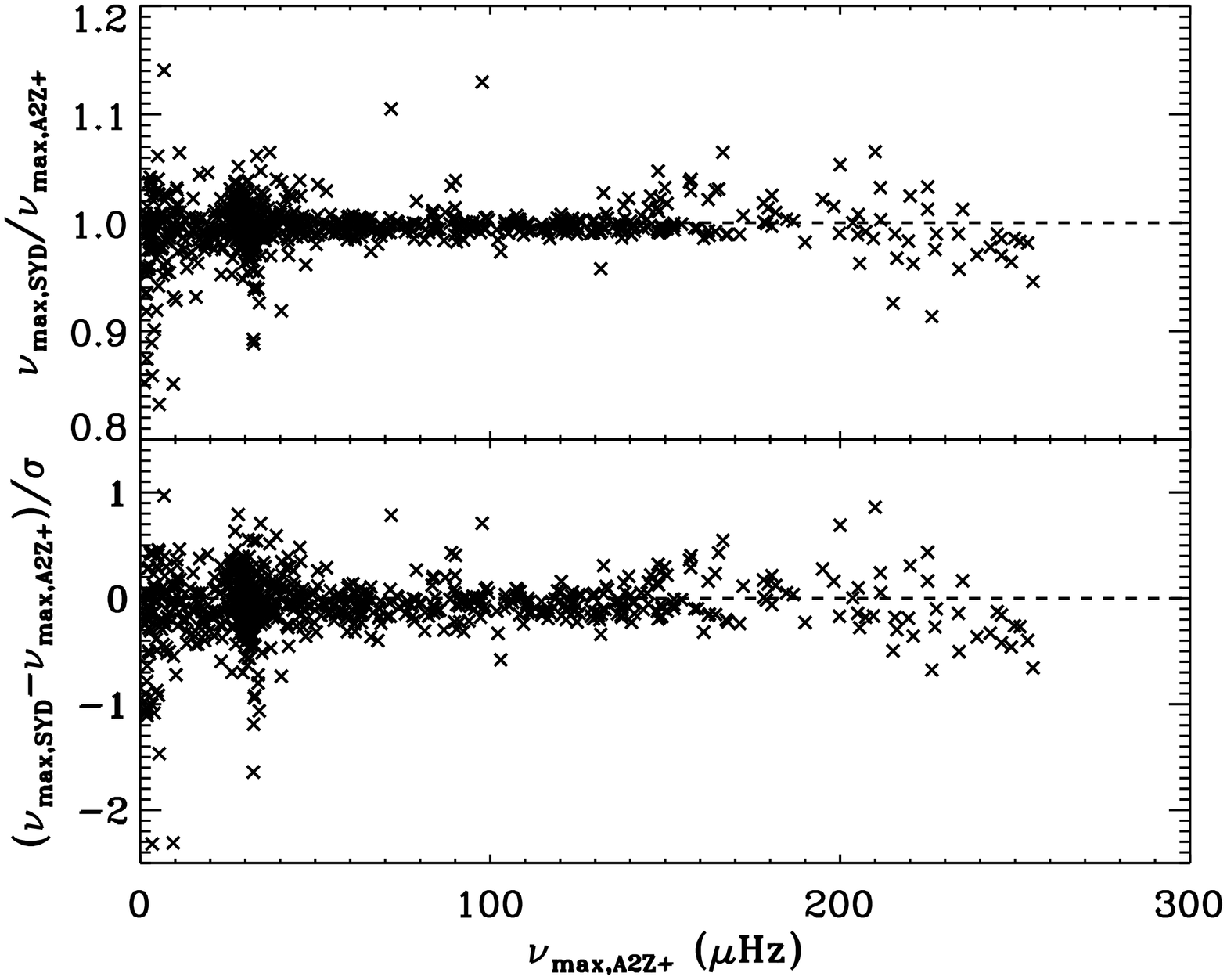}
\caption{Difference in the mean large separation (left panel) and the frequency of maximum power (right panel) between the two pipelines A2Z+ and SYD as a function of the A2Z+ values (top panel) and in units of $\sigma$ computed as the quadratic sum of the uncertainties from A2Z+ and SYD (bottom panel).}
\label{comp_a2zsyd}
\end{center}
\end{figure}

The $\dnu$ and $\numax$ comparison for 748 stars is shown in Figure~\ref{comp_a2zsyd}. For 97.7\% of the stars, the agreement in $\dnu$ is better than 1\,$\sigma$. Concerning $\numax$, both pipelines agree within 1\,$\sigma$ for $\sim$\,98.7\% of the stars. The relative difference in $\dnu$ (resp. $\numax$) is lower than 3\% for 91\% (resp. 85\%) of the stars. The disagreement is larger in two regions: at very low frequency ($\numax <10\mu$Hz and $\dnu < 2.5\mu$Hz) and around $\numax \sim$ 30\,$\mu$Hz and $\dnu \sim$ 4\,$\mu$Hz.

At very low frequency, due to limited resolution and the lower number of modes observed,  it can be more difficult to fit a Gaussian to the excess power (method used by A2Z+ to estimate $\numax$). Moreover, the computation of $\numax$ is obtained after subtracting the background. As a consequence, if the background is fitted slightly differently by each method, it can affect the position of the maximum power.  Moreover, in this frequency-range, the modes show stronger non-asymptotic behavior \citep{2014ApJ...788L..10S}. So a slight difference in the number of orders taken to compute the mean large frequency separation and the frequency range can have an impact on the final estimate of $\dnu$. We did a few tests with A2Z+ and found that in some cases it can lead to more than 3\% change in $\dnu$ for low $\numax$.



The second frequency range where the largest disagreements are found corresponds to the clump stars where the mixed modes (resulting from the coupling between the p-mode and g-mode cavities) are clearly present, making the determination of $\Delta \nu$ more difficult to obtain.


There are 18 stars that disagree by more than 1\,$\sigma$ in terms of $\dnu$. For 11 of them the difference is of the order of a few 0.01$\mu$Hz. For the remaining 7 stars the SNR is too low to see if one value is better than the other. In terms of $\numax$, ten stars disagree by more than 1$\sigma$ among which two (KIC 5351659 and 10587397) have a very low SNR. After visual inspection, for six stars (KIC 2422890, 4383163,  5976435, 9589159, 9712670, and 12599753) the A2Z+ value seems to be a better representation of $\numax$ while for two stars (KIC 5036900 and 6525060) the SYD value seems to be closer to the observed $\numax$. 


We note that the uncertainties provided by the two pipelines are quite different.  For $\numax$, SYD has an average uncertainty of $\sim$2\% and A2Z+ around 6\%. For $\dnu$, SYD has an average uncertainty of 0.6\% compared to 3\% for A2Z+. This difference is due to the different methods used. A2Z+ uncertainties are computed based on power centroids methods as described in \citet{2010MNRAS.402.2049H}, which is known to be quite conservative. The uncertainties of the SYD pipeline are computed through Monte-Carlo simulations by perturbing the power spectrum according to a $\chi^2$ distribution with two degrees of freedom, and repeating the $\numax$ and $\dnu$ measurements 500 times per star. The uncertainties are then estimated from the standard deviation of the resulting $\numax$ and $\dnu$ distributions.

\subsection{Stellar fundamental parameters}

\begin{figure}[h!]
\begin{center}
\includegraphics[angle=90,width=8cm]{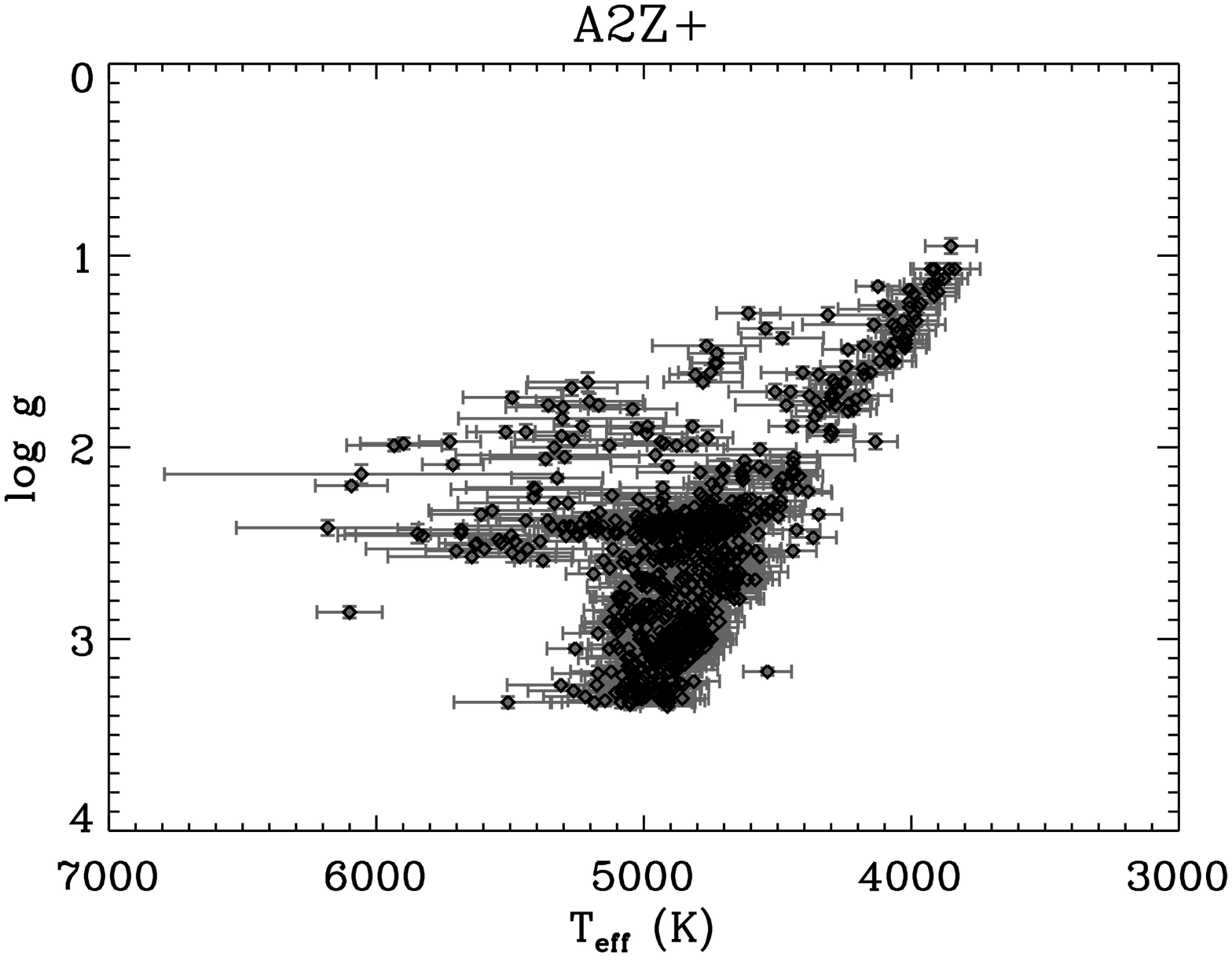}
\includegraphics[angle=90,width=8cm]{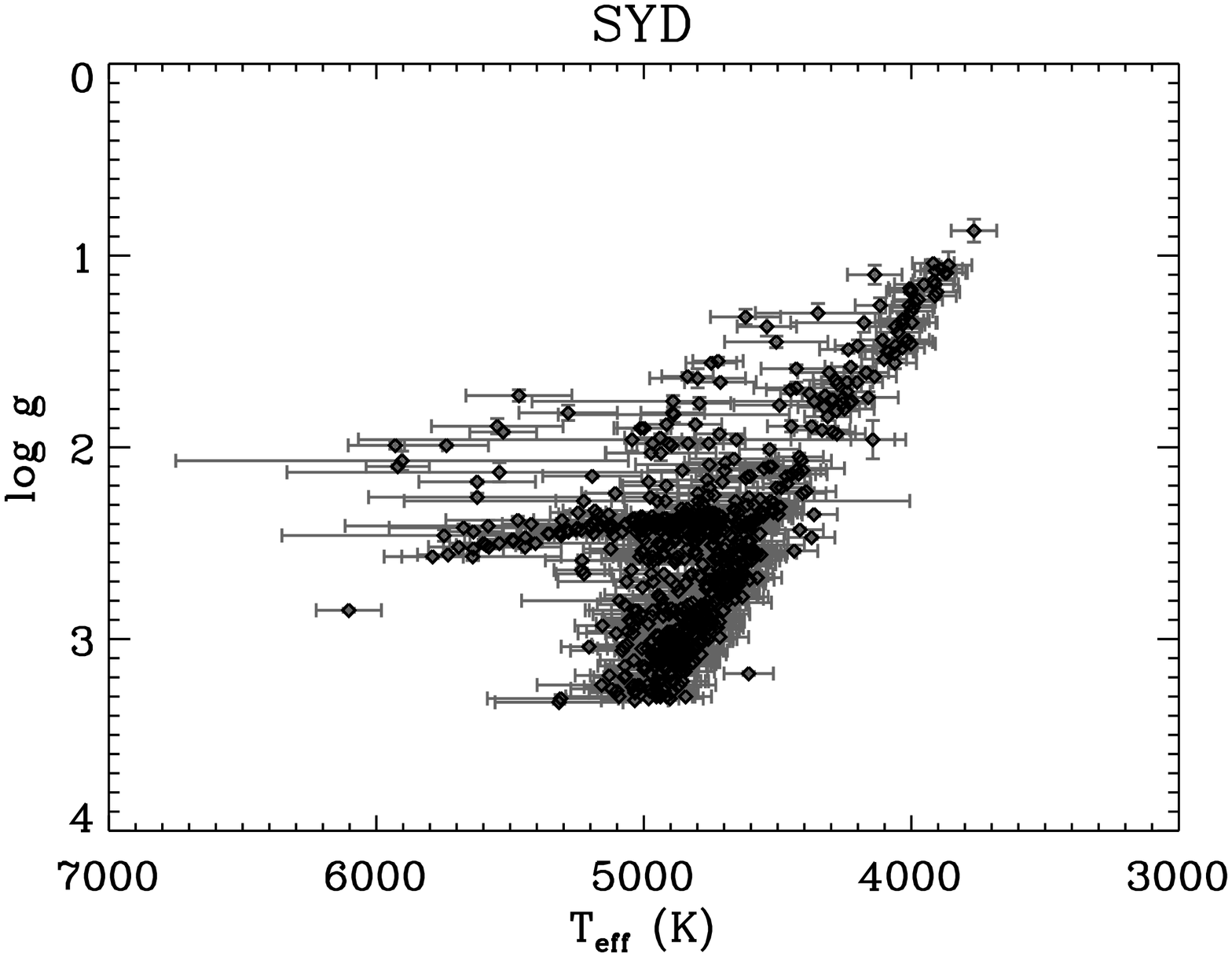}
\caption{HR Diagram obtained for the two pipelines where the effective temperature was computed with the isochrone fitting code (see Section 3.4) and the surface gravities were calculated from the scaling relations.}
\label{comp_HRD_a2zsyd}
\end{center}
\end{figure}

As explained in Section 3.3, we computed the revised effective temperature of these stars by fitting isochrones and using constraints on the colors and the seismic parameters ($\dnu$ and $\numax$). Figure~\ref{comp_HRD_a2zsyd} represents the HR Diagram for the two pipelines with the revised effective temperature and the seismic $\logg$ from scaling relations. We can see that they look very similar even with slightly different seismic parameters.

Figure~\ref{comp_a2zsyd} represents the comparison of the effective temperature and the distance for both pipelines. We note that the stellar parameters obtained from the isochrones fitting agree well within the uncertainties. Since the main contributors for the determination of the effective temperatures are the colors, the difference in the uncertainties on this parameter between A2Z+ and SYD is small. The seismic parameters have more impact on the distance derivation. We note that while the average uncertainty on $\dnu$ (resp. $\numax$) is five times (resp. three times) larger for A2Z+ compared to SYD the mean relative uncertainty on the distance is 1.6 times larger for A2Z+ than for SYD.


\begin{figure}[http]
\begin{center}
\includegraphics[angle=90,width=8cm]{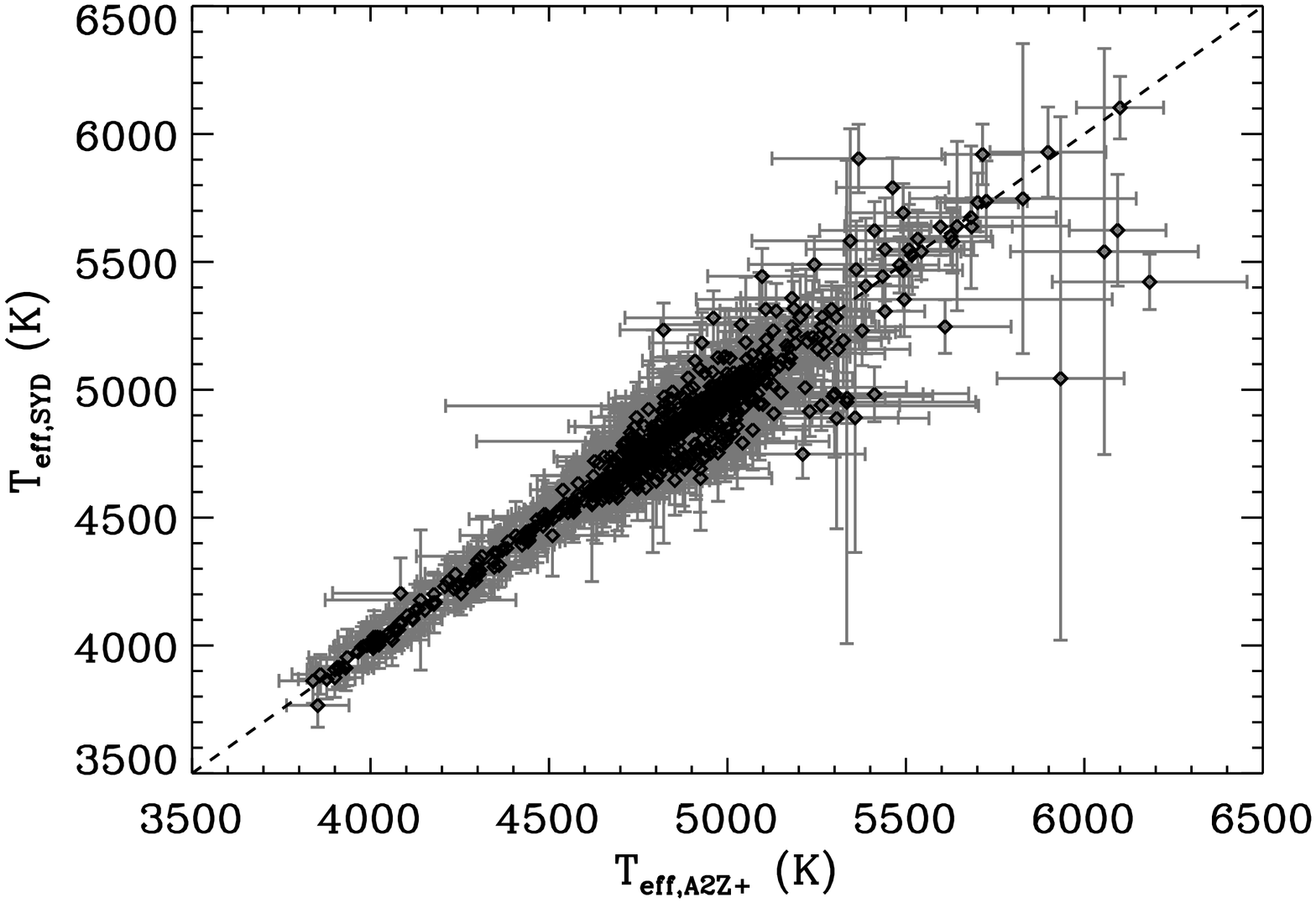}
\includegraphics[angle=90,width=8cm]{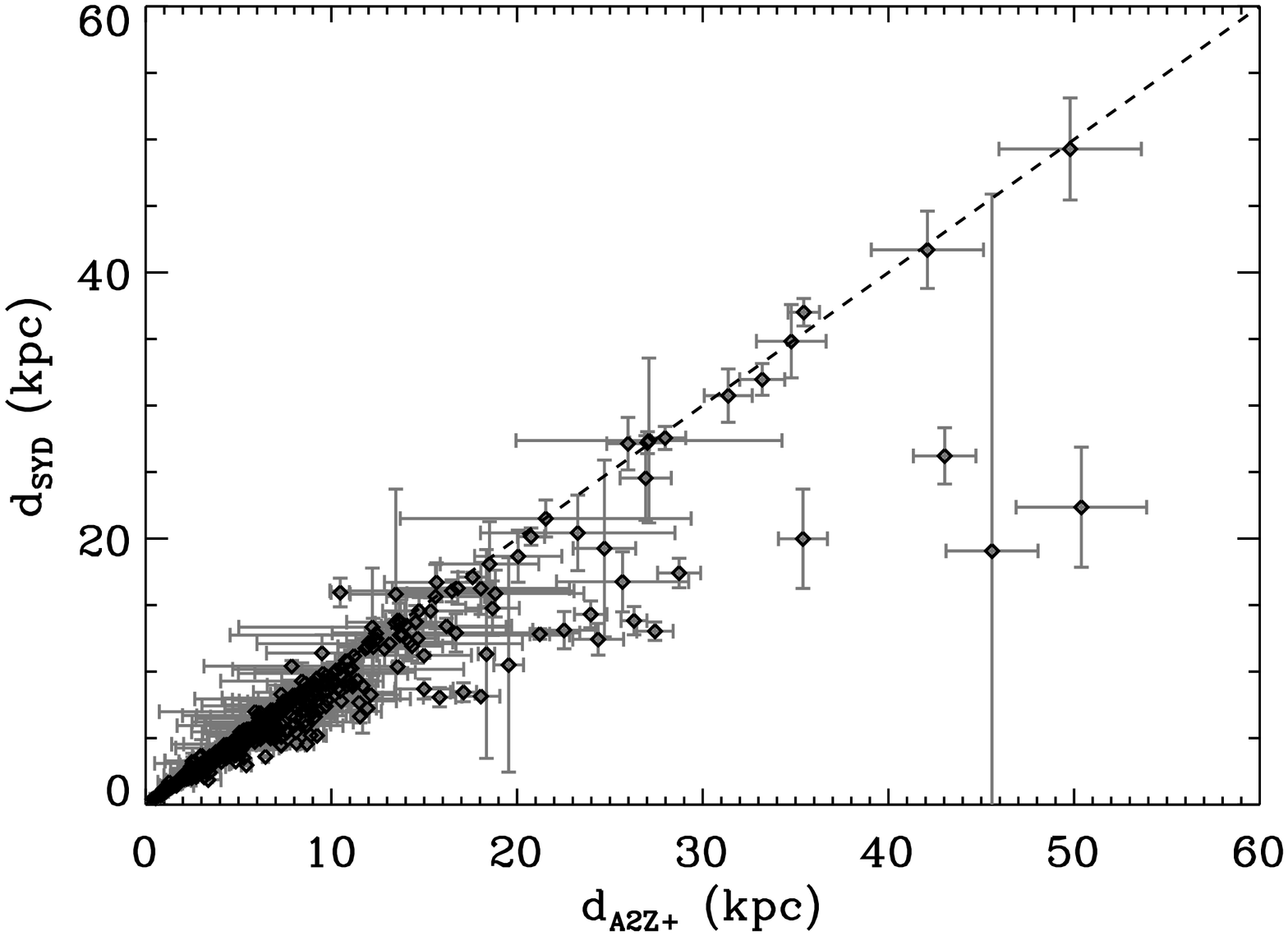}
\caption{Comparison of the effective temperature and distance obtained from the isochrone fitting code for the two pipelines. {  redo d in log scale}}
\label{comp_a2zsyd_teff}
\end{center}
\end{figure}

\begin{figure}[http]
\begin{center}
\includegraphics[angle=90,width=8cm]{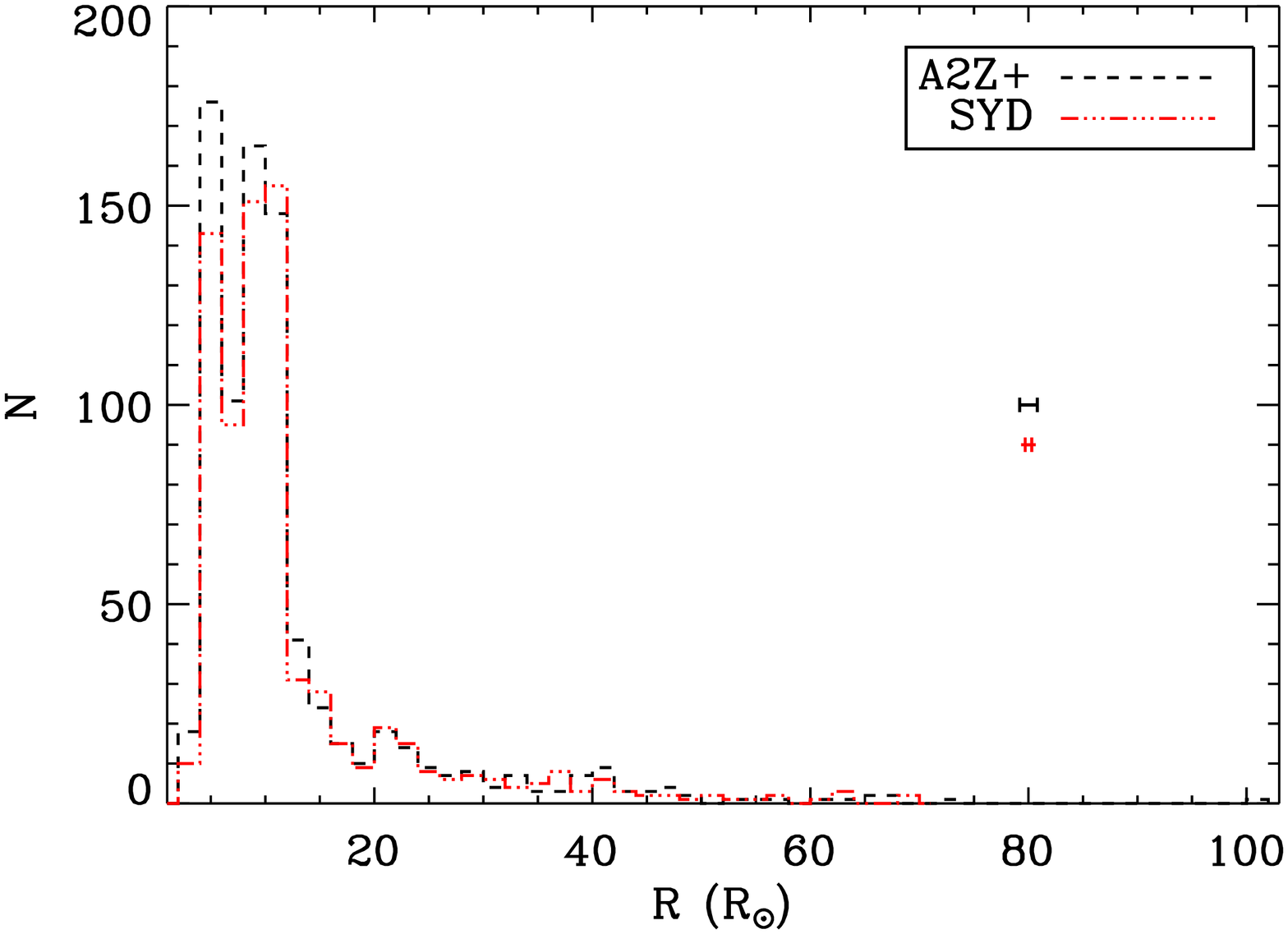}
\includegraphics[angle=90,width=8cm]{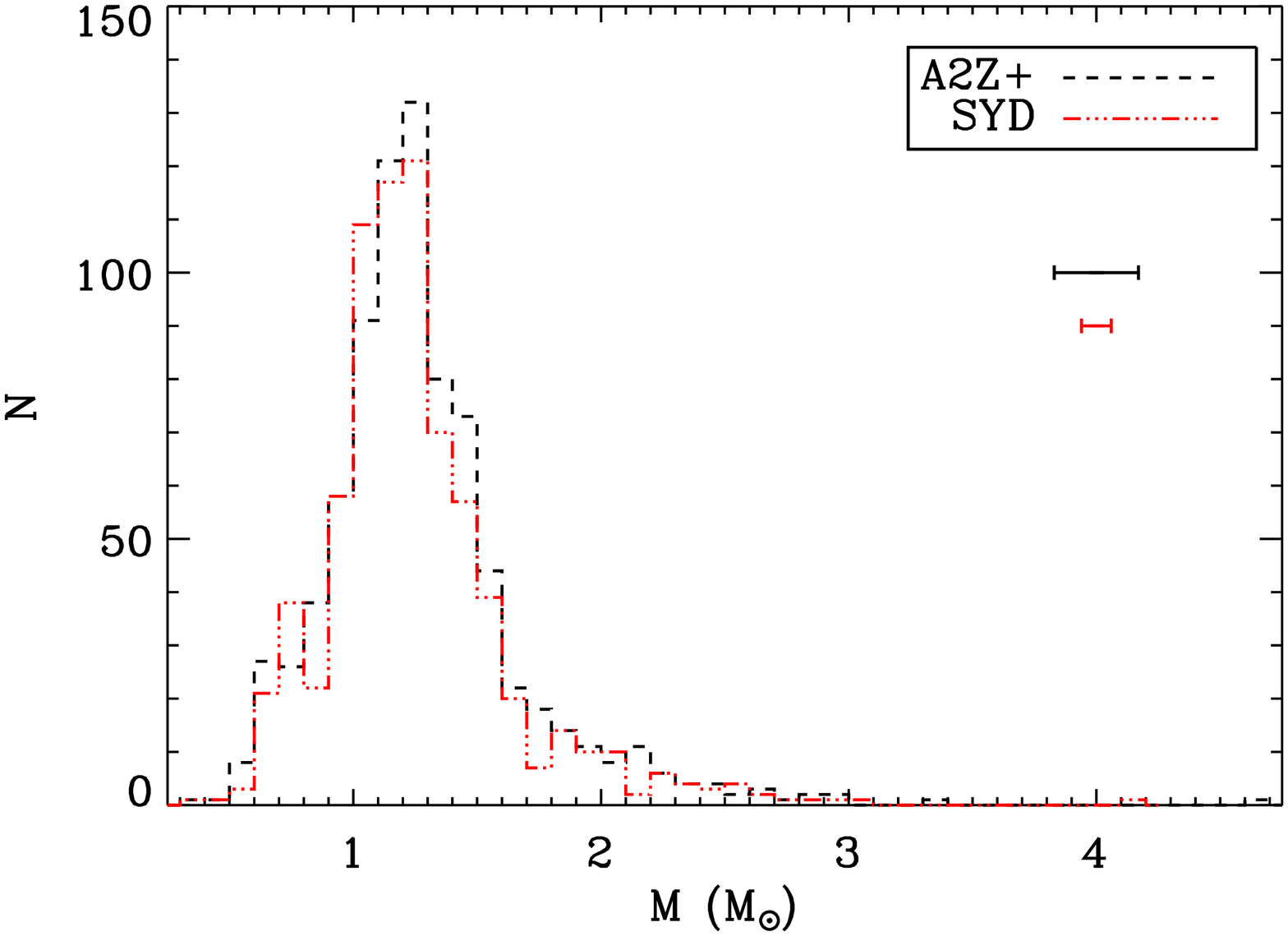}
\caption{Comparison of the radius and mass obtained from the scaling relation for the two pipelines. The error bars on the right hand-side represent the typical uncertainties of each method.}
\label{comp_a2zsyd_MR}
\end{center}
\end{figure}

Finally, we compare the distribution in radius and mass computed with uncorrected scaling relations between A2Z+ and SYD in Figure~\ref{comp_a2zsyd_MR}. We can see that the distributions are quite similar and agree well within the uncertainties. Here we can note the difference in the uncertainties due to the larger uncertainties on the seismic parameters for the A2Z+ pipeline as explained in the previous section.

\end{document}